\documentclass[prd,aps,twocolumn,a4paper,showkeys,nofootinbib,10pt]{revtex4-1}

\usepackage{amsmath}
\usepackage{amsfonts}
\usepackage{amssymb}	
\usepackage{graphicx}
\usepackage{bm}
\usepackage{color}
\usepackage{commath}
\usepackage{hyperref}
\usepackage[T1]{fontenc}
\usepackage{xcolor}
\usepackage{verbatim}
\usepackage{comment}
\usepackage{float}
\usepackage{amsbsy}
\usepackage{makecell}
\usepackage{multirow}

\usepackage[normalem]{ulem}
\allowdisplaybreaks

\def\DALI{{\texttt{TEOBResumS-Dalí}}}

\begin{document}
\title{Direct current memory effects in effective-one-body waveform models}

\author{Elisa Grilli${}^{1,2}$}
\author{Andrea Placidi${}^{1}$}
\author{Simone Albanesi${}^{3,4}$}
\author{Gianluca Grignani${}^{1}$}
\author{Marta Orselli${}^{1,2}$}

\affiliation{${}^1$Dipartimento di Fisica e Geologia, Universit\`a di Perugia,
I.N.F.N. Sezione di Perugia, Via Pascoli, I-06123 Perugia, Italy}

\affiliation{${}^2$Niels Bohr International Academy, Niels Bohr Institute, Copenhagen University,
Blegdamsvej 17, DK-2100 Copenhagen, Denmark}

\affiliation{${}^3$Theoretisch-Physikalisches Institut, Friedrich-Schiller-Universit{\"a}t Jena, 07743, Jena, Germany}
\affiliation{${}^4$INFN sezione di Torino, Torino, 10125, Italy}

% e-mail addresses: one for each author, in the same order as the authors
\email{elisa.grilli@nbi.ku.dk}
\email{andrea.placidi@unipg.it}
\email{simone.albanesi@uni-jena.de}
\email{gianluca.grignani@unipg.it}
\email{orselli@nbi.ku.dk}

\begin{abstract}
The direct current (DC) memory is a non-oscillatory, hereditary component of the gravitational wave (GW) signal that represents one of the most peculiar manifestations of the nonlinear nature of GW emission and propagation. In this work, by transforming the results provided in Ebersold et al. [Phys.Rev.D 100 (2019) 8, 084043] in harmonic coordinates and quasi-Keplerian parametrization, we provide the DC memory in terms of the effective-one-body (EOB) phase-space variables, with a relative accuracy of 2.5PN and in an expansion for small eccentricity up to order six.
Our results are then implemented in \DALI{}, thus 
providing the first EOB model with DC memory contributions. This model is then
used to assess the impact on the waveform and the main features of the DC memory, also addressing its dependence on the eccentricity of the binary system at its formation.
\end{abstract}
\maketitle
%%%%%%%%%%%%%%%%%%%% SECTION %%%%%%%%%%%%%%%%%%%%
\section{Introduction}
\label{sec:intro}   

The observation of the first gravitational wave (GW) signal~\cite{LIGOScientific:2016aoc} by LIGO-Virgo-Kagra (LVK) collaboration~\cite{KAGRA:2021vkt} marked a great leap forward in our understanding of spacetime providing a new way to test the predictions of General Relativity. With the development of next-generation detectors, both space-based, like LISA~\cite{2017arXiv170200786A}, and ground-based, such as Einstein Telescope~\cite{Maggiore:2019uih} and Cosmic Explorer~\cite{Reitze:2019iox,Evans:2021gyd}, we expect GW detections to increase in number, quality, and variety. Reasonably, this will pave the way for investigating aspects of General Relativity that have remained unobserved so far, such as, for example, the so-called GW memory. 
%To have a physical intuition of 
This effect, which is present in all GW signals, 
%it is usual to characterize it as a 
is given by the difference in the baseline of the GW strain, $h(t)$, between early and late times, i.e.
\begin{equation}
    \Delta h_{\rm mem}\, =\, \lim_{t \to +\infty} h(t)-\,\lim_{t\to -\infty}h(t).
\end{equation}
For an ideal detector that is only sensitive to gravitational forces, the GW memory prevents the detector from returning to its initial state after the passage of a GW, causing a difference between the initial and final displacement states.

Although the memory effect causes a change in the proper distance between two free-falling objects, which may seem ideal for detection with an interferometer, due to the non-oscillatory nature of the memory effect and its relatively low frequency and low amplitude profile, it is not expected to play a dominant role in the signals detected by current observations. Nevertheless, numerous studies have explored the detectability of the GW memory, both with current~\cite{Hubner:2019sly, Boersma:2020gxx} and future detectors~\cite{PhysRevD.107.064056, Goncharov:2023woe}, as well as its observational consequences~\cite{PhysRevD.101.104041,PhysRevD.104.123024, Lopez:2023aja}. Indeed, it is likely that it will be possible to detect the GW memory produced by a single merger event between two stellar-mass black holes~\cite{Johnson:2018xly} with the next generation of ground-based detectors, such as Einstein Telescope~\cite{Punturo:2010zz} or Cosmic Explorer~\cite{Evans:2021gyd}. Meanwhile, the future space-based detector LISA~\cite{Audley:2017drz} should be able to measure the GW memory from a single merger between two supermassive black holes~\cite{Favata:2009ii,Islo:2019qht}.

There are two main types of memory effects: the linear memory and the non-linear memory.\footnote{In recent years, linear and nonlinear memory have often been referred to as ordinary and null memory, respectively.} The linear memory arises from a net change in the time-derivatives of some source-multipole moments~\cite{Zeldovich:1974gvh}, and it is relevant for sources such as binary systems on hyperbolic orbits~\cite{1977ApJ-216-610T}, or in the context of supernova explosions~\cite{Sago:2004pn} associated 
gamma-ray-burst jets~\cite{Ott:2008wt}, and asymmetric mass loss due to neutrino emission~\cite{1978ApJ-223-1037E}. 
The non-linear memory, or Christodoulou memory~\cite{PhysRevLett.67.1486,PhysRevD.44.R2945,PhysRevD.45.520}, stems from changes in the radiative multipole moments, and it is sourced by the energy-flux of other emitted GWs. This effect is present in all GW sources and offers a unique and interesting manifestation of the non-linear nature of General Relativity. 

The non-linear memory has already been studied analytically within the Post-Newtonian (PN) expansion during the inspiral phase for several types of binary systems~\cite{Kidder:2007rt, Favata:2008yd, Favata:2011qi, Bieri:2011zb, DeVittori:2014nza}.
Nonetheless, the memory effect is primarily accumulated during the merger phase of a binary black hole (BBH) coalescence, where most of the system's energy and angular momentum is emitted through GWs~\cite{Favata:2008ti}. 
%Numerical Relativity (NR) simulations produce GWs with memory effects, giving a good estimate of the memory through the merger phase. 

Only recently~\cite{Mitman:2020pbt,Yoo:2023spi,Mitman:2024uss}, 
%the memory contribution has been implemented in numerical relativity (NR) simulations in the case of quasi-circular orbits. 
the memory contribution has been extracted from numerical relativity (NR) simulations in the case of quasi-circular orbits. 
This renewed interest that memory effects has received by the scientific community is also sparked by their intriguing connection with soft theorems~\cite{Strominger:2014pwa,Pasterski:2015tva,Compere:2019odm,He:2014laa, Haco:2018ske, Hawking:2016msc, Compere:2019gft,Nichols:2017rqr}. In particular, the memory effects are tied to the so-called Bondi-Metzner-Sachs (BMS) group~\cite{Bondi1962GravitationalWI,Sachs1962GravitationalWI}, which is an infinite-dimensional generalization of the Poincar\'e group. From the BMS  transformations (supertranslation, superrotations and superboosts) it is possible to identify three different types of memory contributions: the displacement memory (or normal memory), which is related to supertranslations and appears as a displacement in the strain; the spin memory~\cite{Nichols:2017rqr,Grant:2022bla}, related to superrotations; the center of mass memory~\cite{Nichols:2018qac}, related to superboosts.\footnote{These three types of memory contributions contain linear and non-linear memory.}

The increasing number of NR simulations with memory contributions and the potential to detect this effect with next-generation detectors are strong drivers for incorporating memory effects produced by various sources into analytically based waveform models. 

%In particular, the non-linear memory contributions can be distinguished in the so called ``oscillatory" memory and ``non-oscillatory" memory terms, the latter also known as direct current (DC) memory. 

A first step toward including memory contributions in analytical waveform models was taken in~\cite{Placidi:2023ofj}, where the oscillatory part of the memory contributions to the waveform~\cite{Ebersold:2019kdc} was implemented in \DALI{}. \cite{Nagar:2024dzj, Nagar:2024oyk}, the eccentric orbit iteration of the \texttt{TEOB} series of effective one-body (EOB) models \cite{Nagar:2021gss,Albanesi:2021rby, Nagar:2024dzj, Nagar:2024oyk}. 
The EOB approach~\cite{Buonanno:1998gg,Buonanno:2000ef,Damour:2000we,Damour:2015isa} enables a unified and theoretically complete description of the full coalescence process of a compact binary, efficiently combining analytical results and NR calibrated components. This results in fast and accurate waveform models that are now an integral part of the detection and data analysis pipelines used in GW astronomy.

Intending to continue in this direction, the purpose of the present work is to focus on the memory component that is still missing in EOB models, the so-called direct current (DC) memory, derived for eccentric orbits in Ref.~\cite{Ebersold:2019kdc}. In particular, we will recast the available results for this DC memory in EOB coordinates, discuss their implementation in \DALI{}, and finally use the so-obtained model to get some insights on the DC memory and its impact on the waveform.  

Accordingly, this paper is organized as follows. In Sec.~\ref{sec:dc_mem_in_hlm}, we briefly review where and how the DC memory appears in the gravitational waveform when the latter is decomposed into spherical harmonic modes. In Sec.~\ref{sec:from_QK_parameters_to_EOB_coord} we derive the 2PN-accurate transformation from the quasi-Keplerian (QK) parametrization of the harmonic coordinates, employed in Ref.~\cite{Ebersold:2019kdc}, to the EOB phase-space variables. These transformations are used in Sec.~\ref{sec:: DCmemEOB} to compute the DC memory contributions in EOB coordinates, with an overall accuracy of 2.5PN~\footnote{To clarify our conventions, PN orders at the level of the waveform are always counted from the dominant $\mathcal{O}(c^{-4})$ quadrupolar contribution.} and under a small eccentricity expansion up to order six.
In Sec.~\ref{sec: properties_DC_mem}, 
we implement these results in the EOB model \DALI{}
and we study the phenomenology of the DC contribution for different eccentric configurations. The impact of
the (2,0) mode on the GW strain is also discussed.
Finally, we make our concluding remarks in Sec.~\ref{sec:conc}.

%%%%%%%%%%%%%%%%%%%% SECTION %%%%%%%%%%%%%%%%%%%%
\section{DC memory contributions in the spherical modes of the waveform}
\label{sec:dc_mem_in_hlm} 

In our analysis, the GW source is a non-spinning black hole binary. Therefore, the whole two-body evolution is planar, with no spin-induced orbital precession.
As usual, we can introduce some useful parameters depending on the masses $m_{1,2}$ of the constituent black holes: the total mass $M\equiv m_1 + m_2$, the reduced mass $\mu\equiv m_1 m_2/M$, and the symmetric mass ratio $\nu\equiv \mu/M$.
%Using the conventions and the notations of\cite{Mishra:2015bqa,Boetzel:2019nfw,Ebersold:2019kdc}, 
For the strain waveform, we consider the standard decomposition into the spherical harmonic modes $h_{\ell m}$ via
\begin{equation}
    h_+ - ih_\times = \sum_{\ell =2}^{\infty}\sum_{m=-\ell}^{\ell} h_{\ell m}\,_{-2}Y^{\ell m}(\Theta, \Phi), 
\end{equation}
where $\,_{-2}Y^{\ell m}(\Theta, \Phi)$ are the spin-weighted spherical harmonics with spin weight $-2$. The spherical harmonic multipoles $h_{\ell m}$ are defined in terms of the radiative mass and current multipoles of the binary, respectively denoted as  $U_{\ell m}$ and $V_{\ell m}$, as follows 
%(we add $c$ to see the power with which it enters, even if $c=1$)
%
\begin{equation}
    h_{\ell m}= - \frac{1}{\sqrt{2} R c^{\ell +2}}\bigg( U_{\ell m}-\frac{i}{c} V_{\ell m}\bigg)\,.
\end{equation}
In the specific case of non-precessing binaries the above relation further simplifies and one finds~\cite{Faye:2012we}
\begin{align}
    h_{\ell m}&=-\frac{U_{\ell m}}{\sqrt{2}Rc^{\ell +2}}~,&& \text{if }\, \ell +m\, \text{ is even,} \\
     h_{\ell m}&=i\, \frac{V_{\ell m}}{\sqrt{2}Rc^{\ell +3}}~,&&\text{if }\, \ell +m\, \text{ is odd.}
\end{align}
While the radiative multipoles $U_{\ell m}$ and $V_{\ell m}$ have been obtained with different techniques~\cite{RevModPhys.52.299, Blanchet:1995fr, PhysRevD.65.064005} and for different classes of compact binaries~\cite{Kidder:2007rt,Arun:2004ff,Arun:2007sg}, in the following, we will refer to the results obtained within the PN-matched multipolar post-Minkowskian approach~\cite{Blanchet:2013haa,Blanchet:1992br, Blanchet:1997ji,Blanchet:1997jj}, and more specifically to their eccentricity-dependent generalization for non-circularized binaries~\cite{Mishra:2015bqa,Boetzel:2019nfw,Ebersold:2019kdc}.

 Notably, in the final expression of these radiative multipoles, one can always identify two types of contributions: instantaneous and hereditary. The former refers to the components of the gravitational wave determined by the source at a specific retarded time, while the latter depends, via time integrals, on the source's history before that retarded time. Furthermore, hereditary terms can be split into two sub-categories based on their specific dependence on the history of the source. We have, in particular, tail terms, which are generally suppressed as one goes towards the remote past of the source and correspond to the back-scattering of GWs off the mass monopole of the source, and memory terms that instead do not fall off as much in the remote past and can be associated to GWs sourced by other, previously emitted GWs. 
 
 At the level of the radiative multipole, this results in a general structure of the type
\begin{align}
    U_{\ell m}&=\, U_{\ell m}^{\rm inst}\, +U_{\ell m}^{\rm tail}\, +U_{\ell m}^{\rm mem}\, +\delta U_{\ell m}, \\ 
    V_{\ell m}&=\, V_{\ell m}^{\rm inst}\, +V_{\ell m}^{\rm tail}\, +\delta V_{\ell m}
\end{align}
where $\delta U_{\ell m}$ and $\delta V_{\ell m}$ denote higher-order terms due to the nonlinear nature of the gravitational interaction and are given by combinations of the hereditary effects described above, such as the tail-of-tail~\cite{Blanchet:1997jj}, the tail-of-memory~\cite{Trestini:2023wwg}, and so on.

Different hereditary effects appear in the PN expansion of the radiative multipoles at different PN orders, with the first tail contributions entering at 1.5PN order and the memory ones at 2.5PN order. For what concerns the memory contributions, however, it is important to take into account the accumulation effects that arise in the computation of the relative time integrals whenever the oscillations of their integrands (as functions of time) are slower than the orbital time scale. While this is relevant for all spherical modes, due to the presence of terms oscillating on the periastron precession time scale, which lead to oscillatory memory contributions already at the 1.5PN level, the spherical modes with $m=0$ are particularly affected by this accumulation effect, since they are the only ones presenting also a purely non-oscillatory memory component, the DC memory we are focused on in this work. More specifically, the DC memory component of each $m=0$ spherical mode comes with an enhancing factor $c^5$ that lowers its PN order by two and a half. Consequently, the DC memory enters the waveform at the leading Newtonian order, and for each mode with $\ell>2$ it appears one full PN order before the leading instantaneous term.\footnote{For instance, the DC memory component of $h_{40}$ starts at the Newtonian order, whereas its instantaneous part starts at 1PN.}

In general, rather than computing $U_{\ell m}^{\rm mem}$ directly from the corresponding multipolar post-Minkowskian expression, it is more convenient to use the result from Refs.~\cite{Blanchet:1992br, Blanchet:1986dk} and derive it through the following formula
\begin{align}\label{eq:Ulmem}
	U_{\ell m}^{\rm mem} &= \frac{32\pi}{c^{2-\ell}} \sqrt{\frac{(\ell -2)!}{2(\ell +2)!}}
		\int_{-\infty}^{T_R} d t \int d \Omega \frac{d^2 E_{\rm GW}}{d t d \Omega} \bar Y_{\ell
		m}(\Omega),
\end{align}
where $T_R$ is the retarded time at which the waveform is evaluated, $Y_{\ell
		m}(\Omega)$ are the standard spherical harmonics, and 
\begin{align} 
\label{eq:energy_flux}
	\frac{d^2 E_{\rm GW}}{d t d \Omega} &\equiv \frac{c^3 R^2}{16 \pi G} \Bigl(
		\dot{h}_+^2 + \dot{h}_\times^2 \Bigr)
\end{align}
is the GW energy flux.

This is precisely the strategy adopted in Ref.~\cite{Ebersold:2019kdc} to get their results for the DC memory
%\mar{to compute...} \eg{to compute the non-linear memory contributions to the GW amplitudes for compact binaries in eccentric orbits}
while using the QK parametrization and considering an expansion around zero on the time eccentricity $e_t$, both introduced to make manageable the evaluation of the time integral in Eq.~\eqref{eq:Ulmem}. Under these simplifying conditions, and by introducing the frequency parameter $x \equiv (GM\Omega)^{2/3}$, which is related to the harmonic orbital frequency $\Omega = \dot{\varphi}_h$, the master integrals required to compute the DC memory are all of the following form:
  \begin{equation}
  \label{eq:memoryIntegral}
           U^{\rm DC}_{pq} = \int_{-\infty}^{T_R} dt\, x^p(t)\, e_t^q(t),
       \end{equation}
where the half-integer parameter $p$ is determined by the PN order of the given contribution, with the leading Newtonian  order corresponding to $p=5$, and the integer parameter $q$ is relative to its eccentricity order, ranging from $0$, the quasi-circular limit, to six, the highest eccentricity order considered.
 
The evaluation of the integral \eqref{eq:memoryIntegral} is actually performed by changing the integration variable to $e_t$,
i.e. considering
\begin{align}
\label{eq:memoryIntegral_e}
	U^{\rm DC}_{pq} = \int_{e_i}^{e_t(T_R)} d e_t\,\Bigl(\frac{d e_t}{d t}\Bigr)^{-1} x^p(e_t)
	\, e_t^q ,
\end{align}
and then by inserting the time evolution equations for $x$ and $e_t$ (and the ensuing relation between the two, see Appendix B of Ref.~\cite{Ebersold:2019kdc}). Here we highlight the appearance of the parameter $e_i\equiv e_t(-\infty)$, which can be physically interpreted as the eccentricity of the binary system at the time of its formation.

As we will see, converting these DC memory contributions into EOB coordinates and implementing them in the EOB model \DALI{}, which will be discussed in the following sections, will also provide insight into how much the DC memory depends on the value of $e_i$.

\begin{comment}
that at leading order in the PN and eccentricity expansion is in the following form
%
\begin{equation}
\label{eq:relationxe}
    x(e)\, =\, x_0\, \Biggl(\frac{e_{t0}}{e_t}\Biggr)^{12/19},
\end{equation}
where $x_0$ is the value of the Newtonian  parameter at some reference eccentricity $e_0$. 
\end{comment}
%%%%%%%%%%%%%%%%%%%%%%SUBSECTION%%%%%%%%%%%%%%%%%%%%%%%%%
\section{From quasi-Keplerian orbital parameters to EOB coordinates}
\label{sec:from_QK_parameters_to_EOB_coord}

The DC memory contributions for generic non-precessing orbits have been derived in Ref.~\cite{Ebersold:2019kdc} by expressing via the QK parametrization~\cite{1985AIHPA..43..107D,Memmesheimer:2004cv} the harmonic polar variables $(r_h, \varphi_h, \Dot{r}_h, \Dot{\varphi}_h)$. Such parametrization is a PN generalization of the more familiar Keplerian parametrization, which similarly rewrites each coordinate in terms of several orbital parameters specific to the description of elliptic orbits. The results provided in~\cite{Ebersold:2019kdc} are thus ultimately given in terms of these QK parameters.

We aim to recast these results for the DC memory contributions, up to the 2.5PN order, in (rescaled) EOB phase-space coordinates. These are the relative separation in the center of mass frame $r=R/M$, the orbital phase $\varphi$, and the two momenta conjugated to them: the radial momentum $p_r=P_R/\mu$ and the angular momentum $p_\varphi=P_\varphi/\mu M$. 

We do this in two steps. First, we go back from the QK parameters to the harmonic coordinates $(r_h, \varphi_h, \Dot{r}_h, \Dot{\varphi}_h)$ that they parameterize, essentially inverting the equations that make up the QK parametrization; then, we replace the harmonic coordinates with the EOB phase space variables $(r, \varphi, p_r, p_\varphi)$ by exploiting the associated coordinate transformations, given, e.g.,~in Eqs.~(5)-(8) of Ref.~\cite{Placidi:2021rkh} at 2PN accuracy.

Since the DC memory we are addressing here enters at Newtonian accuracy, the transformation between QK parameters and EOB variables must be computed with a higher post-Newtonian (PN) accuracy than was required in Ref.~\cite{Placidi:2023ofj}, where the hereditary contributions, such as the oscillatory memory terms, first appeared at the 1.5PN level. Additionally, it is important to note that the relations between the QK parameters and the EOB variables do not involve half-integer PN orders. As a result, to achieve our target accuracy of 2.5PN, only 2PN-accurate transformations are necessary.

At 2PN accuracy, the QK parametrization of harmonic relative separation and orbital phase is given by 
\begin{subequations}
\begin{align}
    r_h &= a\, (1\, -e_r\, \cos{u}), \\ 
\label{eq::phi}
    \varphi_h-\varphi^0_h &=(1\,+ k)\mathit{v}\, + \mathit{f}_\varphi \, \sin{2 \mathit{v}}\, +\mathit{g}_\varphi\, \sin{3 \mathit{v}},
\end{align}
\end{subequations}
where $a$ is the semi-major axis, $u$ is the eccentric anomaly, $k$ is the periastron advance per radial period, $(\mathit{f}_\varphi,\mathit{g}_\varphi)$ are known orbital functions, 
$\varphi^0_h$ is the phase at the first passage of the periastron,
and
\begin{equation}
    \mathit{v}\, =\, 2 \arctan \Biggl[ \Biggl(\frac{1\, +e_\varphi}{1\, -e_\varphi}\Biggr)^{1/2} \tan\frac{u}{2} \Biggr]
\end{equation}
is the true anomaly. Moreover, there are three types of eccentricity: the time eccentricity $e_t$ we already encountered in Sec.~\ref{sec:dc_mem_in_hlm}, the radial eccentricity $e_r$ and the angular eccentricity $e_\varphi$, all three reduce to the Newtonian  eccentricity in the Newtonian  limit. 

For our translation procedure, what we actually need is the 2PN-accurate transformations connecting $x$ and $e_t$ to the EOB coordinates  $(r, \varphi, p_r, p_\varphi)$. Their PN profile reads
\begin{align}
   x=& \frac{1}{c^2}\, x_{\rm N}+\frac{1}{c^4}\, x_{\rm 1PN}+\frac{1}{c^6}\, x_{\rm 2PN}, \\
    e_t=&\, e_{\rm N}\, +\frac{1}{c^2}\, e_{\rm 1PN}\, +\frac{1}{c^4}\, e_{\rm 2PN},
\end{align}
with the leading Newtonian order given by
\begin{align}
x_{\rm N}=&\, -p_r^2\, +u\Bigl(\, 2\, -p_\varphi^2\, u\Bigr), \\
\label{eq:eN}
e_{\rm N}=&\, \sqrt{p_r^2\, p_\varphi^2\, +\Bigl(1\, -p_\varphi^2\, u\Bigr)^2}.
\end{align}

As the 1PN and 2PN coefficient of $e$ and $x$ are quite lengthy, we show their explicit expression in App.~\ref{App:transformation_QK_to_EOB_2PN}. %Mind moreover that the full transformation is also provided in Mathematica form in the Supplementary Material.

Following the conventions of Ref.~\cite{Chiaramello:2020ehz, Nagar:2021gss}, we furthermore rewrite $p_r$ and $p_{\varphi}$ in terms of $(p_{r_*},\Dot{p}_{r_*})$, the momentum conjugated to the tortoise coordinate $r_*=\int\, dr \sqrt{D}/A \, r$ and its time derivative. Here $A$ and $D$ are the effective potentials entering the EOB effective metric.\footnote{More precisely, $D$ is defined as the product $D\equiv AB$ of the two potentials $A$ and $B$, respectively encoding the $dt^2$ and $dr^2$ components of the effective metric in spherical coordinates.} For reference, their 3PN expressions read~\cite{Damour:2000we}
\begin{subequations}
\label{eq:3PN_potentials}
  \begin{align}
          A_{\rm 3PN}=& 1- 2 u+ 2\nu u^3+ \nu \bigg( \frac{94}{3}- \frac{41}{32}\pi^2 \bigg) u^4, \\ 
          D_{\rm 3PN}=& 1- 6\nu u^2+ u^3 \big(-52\nu+ 6\nu^2 \big).
  \end{align}  
\end{subequations}
The main advantage of introducing $p_{r_*}$ is the better behavior it has compared to $p_r$ as the evolution of the binary approaches merger. At the same time, by also replacing $p_\varphi$ with $\Dot{p}_{r_*}$, it is easier to count the eccentricity order of each contribution, according to the simple rule
\begin{equation}
p_{r_*}^n \dot p_{r_*}^m \to (n+m)\text{th eccentricity order}.
\end{equation}
This is particularly important in cases like ours, where the expressions we are dealing with are obtained within an expansion for small eccentricities, and it allows us to remove any spurious contribution that appears beyond the sixth eccentricity order, the highest included in the original result of Ref.~\cite{Ebersold:2019kdc}. 

To rewrite $p_r$ in terms of $p_{r_*}$ we rely on the usual formula
\begin{equation}
\label{eq:prstar}
    p_r = \frac{\sqrt{D}}{A} p_{r_*},
\end{equation}
where the 3PN iterations of the potentials, Eqs.~\eqref{eq:3PN_potentials}, are more than enough for our target 2.5PN accuracy.
By solving perturbatively the PN-expanded equation of motion of $\dot p_r$ for $p_\varphi$, while using the relation between $\dot  p_r$ and  $\Dot{p}_{r_*}$ that follows from Eq.~\eqref{eq:prstar}, i.e.
\begin{widetext}
\begin{align}
\label{prstarDot}
    \Dot{p}_r &= \Dot{p}_{r_*} +\frac{2 u}{c^2} \big( \Dot{p}_{r_*}- p_{r_*}^2 u \big)+\frac{u}{c^4} \Big[ \Dot{p}_{r_*} u (4 -3\nu)-5 p_{r_*}^2 u^2(1-\nu)+\Dot{p}_{r_*}p_{r_*}^2(1+ \nu)+p_{r_*}^4 u(1+ \nu)\Big]+\mathcal{O}\bigg(\frac{1}{c^6}\bigg),
\end{align}
we can then rewrite $p_\varphi$ in terms of $\Dot{p}_{r_*}$ as
\begin{align}
\notag
        &p_\varphi = \sqrt{\frac{\Dot{p}_{r_*} +u^2}{u^3}} + \frac{1}{c^2}\Biggl\{ \frac{6 u^4 +\Dot{p}_{r_*}^2(1+\nu) +\Dot{p}_{r_*} u \big[u (9-\nu)+ p_{r_*}^2 (1+\nu)\big]}{4 u^{5/2} \sqrt{\Dot{p}_{r_*} +u^2}}\Biggr\} \\ \notag
        & \quad + \frac{1}{32 u^{7/2}\big(\Dot{p}_{r_*} +u^2\big)^{3/2}c^4}\Biggl\{ 12 u^8(9 -4\nu) + \Dot{p}_{r_*}^4( 1 +8\nu +\nu^2) -2 \Dot{p}_{r_*} u^4\bigg[p_{r_*}^4(1 -\nu +\nu^2) -2 p_{r_*}^2 u( 8 +6\nu +\nu^2)  \\ \notag 
   &\quad+ u^2 (-153 +57\nu +\nu^2) \bigg] +2 \Dot{p}_{r_*}^3 u \bigg[-p_{r_*}^2 (1 -4\nu +\nu^2) +u (24 +23\nu +2\nu^2)\bigg]+\Dot{p}_{r_*}^2 u^2 \bigg[u^2 (241 -20\nu -3\nu^2)  \\
   &\quad-3p_{r_*}^4 (1 +\nu^2) +2 p_{r_*}^2 u (13 +16\nu +3\nu^2)\bigg]\Biggl\} +\mathcal{O}\bigg(\frac{1}{c^6}\bigg).
\end{align}
\end{widetext}
We emphasize that the next order in this relation would be the 3PN, implying that what is shown above is enough for reaching the 2.5PN accuracy we are after.

%%%%%%%%%%%%%%SECTION%%%%%%%%%%%%%%%%%%%%%%%%%%%%
\section{DC memory in EOB waveform models}
\label{sec:: DCmemEOB}
In this section, we provide the EOB coordinate expression of the DC memory contribution to the $m=0$ spherical modes, and describe how to incorporate it into an EOB waveform model, specifically targeting \DALI{}.

Factoring out the constant term of the leading quadrupolar mode, we write the $m=0$ spherical modes as
 \begin{equation}
 \label{eq:sphericalmodesConvention}
      h_{\ell 0}=\frac{8\, G\, M\, \nu}{c^2\, R}\, \sqrt{\frac{\pi}{5}}\,
      %e^{-i\, m\, \psi}\, 
      H_{\ell 0}
   \end{equation}
where, isolating each different contribution, we have   
\begin{equation}
     H_{\ell 0}=\, H_{\ell 0}^{\rm inst}\, +H_{\ell 0}^{\rm tail}\, +H_{\ell 0}^{\rm mem},
\end{equation}
with the memory contribution that can be further split in an oscillatory and DC part, $H_{\ell 0}^{\rm mem,osc}$, and $H_{\ell 0}^{\rm DC}$ respectively.    
Even though here we are mainly focused on the DC contribution $H_{\ell 0}^{\rm DC}$, it should be noted that all the other contributions to the $m=0$ modes have been previously provided in EOB coordinates only up to the 2PN order \cite{Placidi:2021rkh}. Thus, in App.~\ref{App:2.5PN_h20} we provide the new 2.5PN contributions to  
$H_{2 0}^{\rm inst}$, $H_{2 0}^{\rm tail}$, and $H_{2 0}^{\rm mem,osc}$, along with the post-adiabatic corrections that appear at this PN order when using the QK parametrization. Contributions of this type for the higher-$\ell$ modes can be found in the Supplementary Material.

% The phase $\psi$ is given by 
%  \begin{equation}
%  \label{eq::psi}
%  \psi \, =\, \varphi\, -2\, x^{3/2}\biggl(1\,-\frac{\nu}{2}\, x\biggr) \ln{\frac{\omega }{\omega_0}}.
%   \end{equation}
% where $\omega$ represent the orbital frequency of the         circular orbit~\cite{Blanchet:2008je}. We can incorporate the $\ln{\omega}$ terms into the phase, as they modify the wave amplitude starting from the 1.5 PN order. However, we can neglect these terms and we can use the phase $\varphi$ \eqref{eq::phi}, because the modification of the phase only appears at the 4PN order. This is because the lowest-order phase evolution corresponds to the inverse of the radiation reaction, leading to a typical 2.5 PN order difference between amplitude and phase~\cite{Blanchet:1996pi}.
% So we can absorb into the phase the $\ln{\omega}$-terms, that appear as some modifications of the wave amplitude from the 1.5 PN. However, we can neglet the $\ln{\omega}$-terms in the phase because these appear from the 4PN order, due to the fact that the lowest-order phase evolution is at the inverse of the radiation-reactiom, so as usual there is a difference of 2.5 PN order between amplitude and phase.

Let us now address the DC part of the modes.
From the QK results of Ref.~\cite{Ebersold:2019kdc}, we can obtain the DC memory contributions in EOB coordinates with the help of the transformations discussed above, in Sec.~\ref{sec:from_QK_parameters_to_EOB_coord}. Specifically, due to the presence of multiple powers of the ratio $e_t/e_i$, we found it helpful to first apply the transformations as if this ratio were constant. Then, we reapply the transformations to all powers of the ratio separately, carefully removing any terms that exceed our desired PN and eccentricity accuracy.

Below we show the result we get for $H_{2 0}^{\rm DC}$ in the EOB coordinates $(u, p_{r_*}, \Dot{p}_{r_*})$, with $u=1/r$ and an accuracy of 2.5PN, although limited for brevity to the leading order in eccentricity. Note that similar expressions for the higher-$\ell$ DC contributions are collected in App.~\ref{App: Higher_Modes_Hlm}, while the corresponding complete expressions up to the sixth order in eccentricity are provided in the Supplementary Material. 

Isolating the different PN orders in 
%$H_{20}^{\rm DC_{\rm N} DC}$ 
$H_{20}^{\rm DC}$ as

\begin{widetext}
    \begin{equation}
    H_{20}^{\rm DC}\,=\, -\frac{5\, u}{14 \sqrt{6}}\Biggl(H_{20}^{\rm DC_{\rm N} }\, +u\, H_{20}^{\rm DC_{\rm 1PN} }\,+ u^{3/2}\,H_{20}^{\rm DC_{\rm 1.5PN} }\, +u^2\, H_{20}^{\rm DC_{\rm 2PN} }\,+ u^{5/2}\,H_{20}^{\rm DC_{\rm 2.5PN} }\Biggr),
\end{equation}
and introducing the quantity $Z_p = \sqrt{\Dot{p}_{r_*}^2\, +p_{r_*}^2\, u^3} $,\footnote{This combination of momenta is proportional to the Newtonian eccentricity \eqref{eq:eN} when the latter is rewritten in terms of $p_{r_*}$ and $\dot p_{r_*}$; we have in fact $e_{\rm N}=Z_p/u^2$.} we find

\begin{subequations}
   \begin{align}
     &H_{20}^{\rm DC_{\rm N}}=\, 1\, -\Biggl(\frac{Z_p}{e_i u^2}\Biggr)^{\frac{12}{19}}, \\ \notag 
        &H^{\rm DC_{\rm 1PN}}_{20}=\,-\frac{4075}{4032}\, +\frac{83}{48}\nu\,+\Biggl(\frac{Z_p}{e_i u^2}\Biggr)^{\frac{24}{19}}\Bigl(\frac{145417}{76608}-\,\frac{23931}{76608}\nu\Bigr)+\Biggl(\frac{Z_p}{e_i u^2}\Biggr)^{\frac{12}{19}}\Biggl[\,\frac{4452\nu}{3192}\,-\frac{2833}{3192} \\ 
        &\quad+\frac{84}{3192\,Z_p^4}\Bigl(p_{r_*}^4\, u^6(\,18\,-12\nu)\, -\Dot{p}_{r_*}^4(\,18\, +6\nu)\,-18\,p_{r_*}\, \Dot{p}_{r_*}^2\, u^3\, \nu \Bigr) \Biggr], \\
        &H_{20}^{\rm DC_{\rm 1.5PN} }=\, \frac{377}{228}\pi\, \Biggl(\frac{Z_p}{e_i u^2}\Biggr)^{\frac{12}{19}}\Biggl[1\, -\Biggl(\frac{Z_p}{e_i u^2}\Biggr)^{\frac{18}{19}}\Biggr], \\
        \notag
        &H_{20}^{\rm DC_{\rm 2PN} }=\, -\frac{151877213}{67060224}\, -\frac{627415}{133056}\nu\, +\frac{5497}{3168}\nu^2\,+\Biggl(\frac{Z_p}{e_i u^2}\Biggr)^{\frac{36}{19}}\Biggl[\, -\frac{11654209}{1143648}\nu^2 -\frac{50392977379}{24208740864}\,
        +\frac{764295307}{48033216}\nu  \\ \notag
        & \quad-\frac{50392977379}{24208740864} \Biggr]\, +\Biggl(\frac{Z_p}{e_i u^2}\Biggr)^{\frac{12}{19}}\Biggl[ \frac{358353209}{366799104}\, -\frac{259303}{727776}\nu\, -\frac{1495}{5776}\nu^2\, +\frac{1}{10108\, Z_p^4}\Biggl( -p_{r_*}^2\Dot{p}_{r_*}^2\, u^3(\, 8499 \nu\, -13356\nu^2)  \\ \notag
        & \quad+p_{r_*}^4\, u^6\bigl(8499\,  -19022\nu\, +8904\nu^2 \bigr)\, +p_{r_*}^4\bigl(-8499\, +10523\nu\, +4452\nu^2\bigr) \Biggr)+\frac{1}{722 Z_p^8}\Biggl(\, \Dot{p}_{r_*}^8\bigl( -4314\,+3489\nu  \\ \notag
        & \quad -150\nu^2\bigr)\, +2\, p_{r_*}^8\, u^{12}\bigl(-105\, +1128\nu\, -186\nu^2 \bigr)\, +p_{r_*}^2\, \Dot{p}_{r_*}^6\, u^3\bigl( -15960\,+13659\nu\, -900\nu^2 \bigr)\\ \notag
        & \quad+p_{r_*}^6\, \Dot{p}_{r_*}^2\, u^9\bigl(-7752\, +11193\nu\,- 1344\nu^2\bigr)\, +p_{r_*}^4\, \Dot{p}_{r_*}^4\, u^6\bigl(-19188\, +19107 \nu\, -1722 \nu^2\bigr) \Biggr) \Biggr]  \\ \notag
        & \quad+\Biggl(\frac{Z_p}{e_i u^2}\Biggr)^{\frac{24}{19}}\Biggl[+\frac{411966361}{122266368}\, -\frac{7889159}{727776}\nu\, +\frac{150997}{17328}\nu^2\, -\frac{1}{2911104\, Z_p^4}\Biggl(\, \Dot{p}_{r_*}^4 \bigl(\, -2181255\, +2862655\nu\, +119580\nu^2\bigr)  \\ \notag
        & \quad +p_{r_*}^4\, u^6\bigl(2181255\, -5043910\nu\, +2393160\nu^2\bigr)\, +p_{r_*}^2\, \Dot{p}_{r_*}^2\, u^3\bigl(\, -2181255\nu\, +3589740\nu^2\bigr) \Biggr)\, \\
        &\quad +\frac{1}{153216\, Z_p^2}\Biggl(\,\Dot{p}_{r_*}^2\bigl(-436251\, +572531\nu\, +239316\nu^2\bigr)\, +p_{r_*}^2\, u^3\bigl(\, 436251\, -1008782\nu\, 478632 \nu^2\bigr) \Biggr) \Biggr],  \\ \notag
        &H_{20}^{\rm DC_{\rm 2.5PN} }=\, \pi\, \Biggl\{\frac{253}{336}(\, -1\, +4\nu)\, +\Biggl(\frac{Z_p}{e_i u^2}\Biggr)^{\frac{23}{19}}\Biggl(\, -\frac{424020733}{43666560}\, +\frac{27049187}{3638880}\nu\Biggr)\,+\Biggl(\frac{Z_p}{e_i u^2}\Biggr)^{\frac{24}{19}}\Biggl( \frac{54822209}{8733312}\, -\frac{1074073}{103968}\nu\Biggr)\,  \\ \notag
        &\quad +\Biggl(\frac{Z_p}{e_i u^2}\Biggr)^{\frac{30}{19}}\Biggl[\frac{5340205}{1455652}\, -\frac{99905}{17328}\nu\, +\frac{1}{1444\, Z_p^4}\Biggl(\, 5\, p_{r_*}^4\, u^{6}\bigl(\, -1131\, +754\nu \bigr) \, +5\, \Dot{p}_{r_*}^4\bigl(\, 1131\, +377\nu\bigr) \, +5655\, \Dot{p}_{r_*}^2\, p_{r_*}\, u^{3}\nu \Biggr)\Biggr]\, \\ 
        & \quad +\Biggl(\frac{Z_p}{e_i u^2}\Biggr)^{\frac{12}{19}}\Biggl[\, \frac{3763903}{7277760}\,+\frac{3427243}{606480}\nu\, +\frac{1}{722\, Z_p^4}\Biggl(\, -\Dot{p}_{r_*}^4\bigl(1131\, +377\nu\bigr)\, -1131\, p_{r_*}^2\, \Dot{p}_{r_*}^2\, u^3\nu\,  -p_{r_*}^4\, u^6\bigl(-1131\, +754\nu\bigr)\Biggr) \Biggr] \Biggr\}\,.
   \end{align} 
\end{subequations}
\end{widetext}

Taking the circular limit ($p_{r_*}\to0$ and $\dot p_{r_*}\to0$), we find that our results are consistent with the 3PN DC memory modes computed long ago by Favata in Ref.~\cite{Favata:2008yd}.\footnote{We note that once the circular limit is taken there is no leftover dependence on $e_i$. The same happens when taking the limit $e_t \to 0$ on the original QK expressions.}

We also highlight the presence of the initial eccentricity $e_i$, which originates from the integrals ~\eqref{eq:memoryIntegral_e}. Based on Ref.~\cite{Favata:2011qi}, which investigates the sensitivity of the nonlinear memory to the early history of the binary, it has been shown that the choice of $e_i$ is, at least in principle, arbitrary, as contributions to the memory occurring outside the observation period are undetectable.  Therefore, in the following we will consistently set $e_i=1$, except when assessing the impact of this choice on the memory contribution.

% Moreover, for quasicircular binaries that were initially elliptical, the early-time eccentricity onlu a negligible correction to the overall memory effect
%It can be shown~\cite{Favata:2011qi} that the choice of its value is, at least in principle, arbitrary, so in the following we will always set $e_i=1$, except when assessing the impact of such a choice on the memory contribution.
Note that setting $e_i=1$ is equivalent to assuming that the binary was formed through a parabolic dynamical capture. In astrophysical scenarios, elliptic-like orbits may form via 
hyperbolic captures~\cite{Samsing:2013kua,Rodriguez:2018pss,DallAmico:2023neb}, 
but the binary’s history prior to the first encounter should not affect the DC term.
This is because the DC contribution is related to the time-integral of
$|\dot{h}|^2$,~\footnote{Or, in terms of Bondi-Sachs functions, the squared norm of the News.}
and during hyperbolic motion, the waveform's strain is primarily driven by the black holes’ velocities, and is therefore constant. 
As a result, the contribution to the integral of $|\dot{h}|^2$ during the hyperbolic 
phase is negligible.

Moving forward, let us now outline how to implement this additional analytical information into the state-of-the-art EOB model \DALI{}~\cite{Nagar:2024dzj, Nagar:2024oyk}. To begin, we recall that, as originally pointed out in Ref.~\cite{Placidi:2021rkh}, for the $m=0$ modes it is not beneficial to consider the usual factorization of the waveform introduced in \cite{Damour:2008gu}. This is mainly because the latter entails the factorization of the Newtonian term (i.e., the leading instantaneous term), which is always fully non-circular for the $m=0$ modes, which can introduce spurious poles in the circular limit. Additionally, it has been demonstrated that keeping all time derivatives explicit in the Newtonian terms—rather than reducing them using PN-expanded equations of motion—generally produces more reliable results. This approach has been supported in Refs.~\cite{Chiaramello:2020ehz,Albanesi:2021rby,Placidi:2021rkh,Albanesi:2022xge}.

Therefore, we propose to model the $m=0$ modes according to the simple non-factorized structure
\begin{equation}
    \label{eq:model_hl0}
    h_{\ell 0} = h_{\ell 0}^{\rm N} + h_{\ell 0}^{\rm inst} +  h_{\ell 0}^{\rm tail} + h_{\ell 0}^{\rm post-ad}+ h_{\ell 0}^{\rm mem,osc}+ h_{\ell 0}^{\rm DC},
\end{equation}
with the Newtonian term $h_{\ell 0}^{\rm N}$ kept with explicit time derivatives, such as
\begin{equation}
\label{eq:h20_N}
    h_{20}^{\rm N} = 4\sqrt{\frac{2 \pi}{15}}  \frac{G M \nu}{c^2 R} \Biggl( r \ddot{r} + \dot{r}^2 \Biggr)\,,
\end{equation}
and all the other corrections, including instantaneous, tail, post-adiabatic, oscillatory memory, and DC memory, expressed in terms of $(u, p_{r_*}, \dot{p}_{r_*})$. 

The results we provide in the Supplementary Material for all the $m=0$ modes up to $\ell = 8$, at 2.5PN accuracy and up to the sixth order in eccentricity, are organized according to this prescription. For completeness, we also provide therein the order reduced expressions of the Newtonian terms as functions of $(u, p_{r_*}, \dot{p}_{r_*})$, like all the other contributions.  
%
%%%%%%%%%%%%%%%%%%%% SECTION %%%%%%%%%%%%%%%%%%%%
%\newpage

\section{Inspecting the DC memory within an EOB model}
\label{sec: properties_DC_mem}
\begin{figure}
  \centering 
    \includegraphics[width=0.47\textwidth]{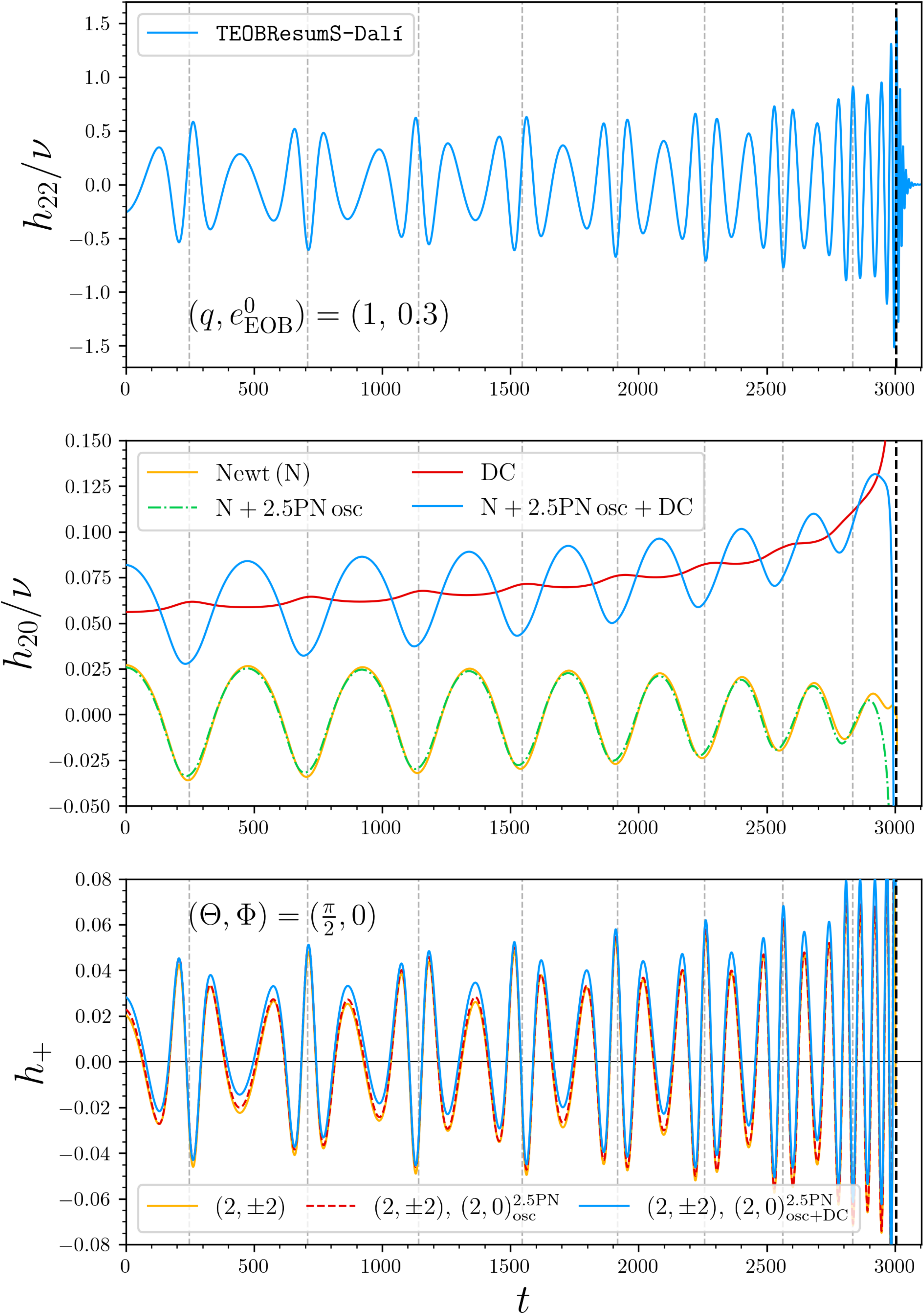}\\
    \caption{
    Quadrupolar waveform from \DALI{} 
    for an equal mass non-spinning binary with initial EOB eccentricity 
    $e_{\rm EOB}^0=0.3$ and orbital-averaged initial frequency
    $\langle f_0 \rangle = 0.005 M^{-1}$. 
    We show the complete (2,2) and (2,0) multipoles with blue lines 
    in the top and middle panel, respectively. In the latter we also 
    report separately the different contributions that enter in the (2,0) mode:
    the Newtonian  generic factor (orange), the waveform up to 2PN (green), and
    the solely DC memory contribution (red). 
    The dotted vertical lines mark the periastra, while the dashed one marks
    the peak of the (2,2) amplitude. 
    In the bottom panel we report the plus polarization for the observational
    direction $(\Theta, \Phi)=(\frac{\pi}{2},0)$ computed up to merger with only the $(2,\pm 2)$ modes
    (orange) and considering also the 2.5PN (2,0) mode, both without and with 
    DC contribution (dashed red and solid blue, respectively).
    }    
    \label{fig:quadrupole_q1_e03}
\end{figure} 

Having clarified how the DC memory contributions can be implemented in the $m=0$ sector of \DALI{}, we are now ready to examine the resulting waveforms. We briefly recall that, once the initial data is specified,
\DALI{} computes the EOB dynamics by solving the Hamiltonian equations of motion for generic orbits
by numerical integration. The resulting dynamics can then be used to evaluate the analytical
expressions for the EOB waveform. 

%Note that we employ the expressions for the (2,0) mode only in the last step, i.e., 
The back-reaction on the dynamics due to the (2,0) mode is not considered; however, 
all other modes with $|m|>0$ and up to $\ell=8$ are included in the 
radiation reaction. For further details we refer to Refs.~\cite{Nagar:2024dzj, Nagar:2024oyk}.

The quadrupolar waveform for an equal mass non-spinning binary is reported in Fig.~\ref{fig:quadrupole_q1_e03}.
The evolution begins at apastron, with an orbital-averaged 
initial frequency of $\langle f_0 \rangle = 0.005\,M^{-1}$
and an initial EOB eccentricity of $e_{\rm EOB}^0 = 0.3$.
It is useful to recall here that the definition of the EOB eccentricity $e_{\rm EOB}$ (and its initial value $e_{\rm EOB}^0$)
is not related to the initial eccentricity $e_i$ mentioned earlier,
which refers to the binary formation.
Instead, the EOB eccentricity is defined, along with the semilatus rectum $p_{\rm EOB}$, 
using the Newtonian relations 
$e_{\rm EOB}\equiv(r_+-r_-)/(r_++r_-)$ and $p_{\rm EOB}\equiv2r_+r_-/(r_++r_-)$,
where $r_\pm$ are the apastron and the periastron of the EOB orbital motion. These radial turning points
can be related to the energy $E$ and the angular momentum $p_\varphi$ of the system by inverting the relations 
$E=V(r_\pm)$, where $V$ is the effective potential associated with the EOB Hamiltonian, defined as
$V(r)\equiv H_{\rm EOB}(r;p_\varphi, p_{r_*}=0)$. 
Finally, we emphasize that, 
while the EOB eccentricity is helpful for generating initial data in a geometrically intuitive way, it is not used during the evolution. However, it can be formally computed up to the time of separatrix crossing, i.e. until $E=V(r)$ has two solutions.

In the top panel of Fig.~\ref{fig:quadrupole_q1_e03} we show the real part of the dominant (2,2) mode,
with dotted vertical lines marking the periastron passages, which correspond to bursts of radiation.
The merger time, defined as the peak of the (2,2) mode amplitude, is marked with a dashed black line. 
In the middle panel, we display the 
(2,0) multipole as derived in this study. The full mode 
is represented by a blue line, while the individual contributions are distinguished by 
different colors. Notably, the oscillatory behavior is primarily captured by 
the Newtonian  correction from Eq.~\eqref{eq:h20_N} (solid orange).  Interestingly,
the significance of the oscillatory part diminishes towards merger, 
due to the strong circularization caused by gravitational wave emission.
When the 2PN instantaneous and tail corrections are added to the Newtonian  term 
(dashed green), the waveform remains largely unchanged during the inspiral and plunge phases. 
However, a nonphysical sharp growth appears just 
before the merger time,  indicating that the PN corrections become unreliable in this regime, as expected.
A similar sharp growth is observed in the DC memory term, shown in red. However, since the 
DC memory is a cumulative effect, it is not immediately clear when the DC term starts to lose reliability.

\begin{figure}
  \centering 
    \includegraphics[width=0.47\textwidth]{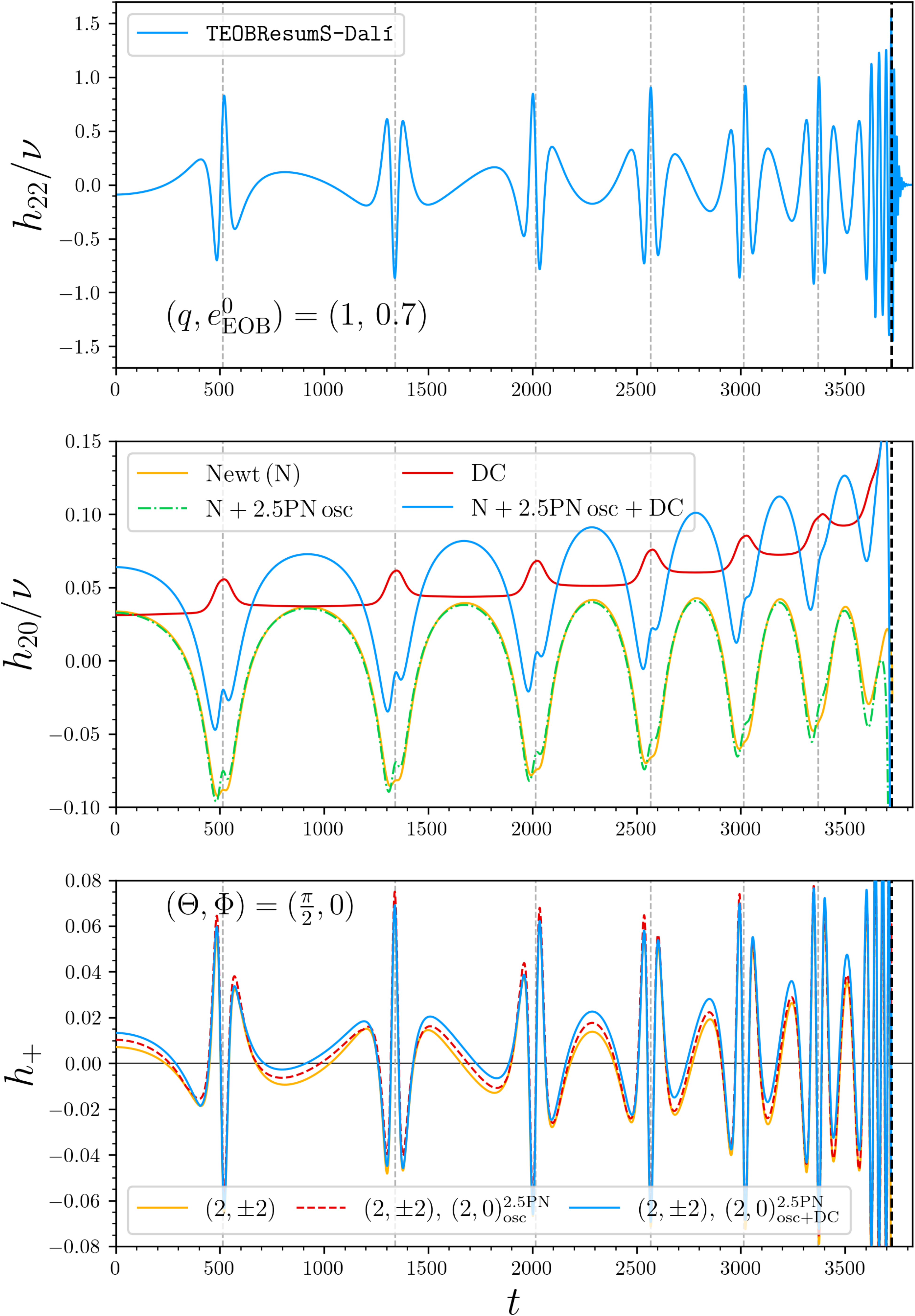}
    \caption{
    Quadrupolar waveform from \DALI{} for an equal mass non-spinning binary
    with initial EOB eccentricity $e_{\rm EOB}^0=0.7$ and $\langle f_0\rangle =0.002 M^{-1}$. Analogous to 
    Fig.~\ref{fig:quadrupole_q1_e03}.
    }    
    \label{fig:quadrupole_q1_e07}
\end{figure}

In the bottom panel we present the plus polarization of the GW
strain computed up to merger for an observer whose line of sight is 
$(\Theta, \Phi)=(\frac{\pi}{2},0)$, which is the direction that 
maximizes the influence of the (2,0) mode.
Note that the (2,0) does not contribute to the cross polarization, 
so that the displacement memory contribution only affects the plus polarization.
We consider three scenarios: 
including only the $(2,\pm 2)$ modes in the strain (solid orange),
incorporating also the oscillatory part of the (2,0) multipole up to 2.5PN (dashed red),
and finally considering the complete (2,0) mode, also with displacement memory (solid blue). 
As can be seen, the contribution of the oscillatory part of the (2,0) mode is relatively
small and does not significantly change the 
strain\footnote{Note that for higher eccentricities, even 
the oscillatory term has a significant impact on the strain, especially 
at apastron. See, e.g., the bottom panel of Fig.~\ref{fig:quadrupole_q1_e07}. However, the DC term
is still dominating over the (2,0) oscillatory part.}. 
However, once that the DC memory contribution
is included, the plus polarization of the strain is clearly distinguishable from the one 
computed with only the $(2, \pm 2)$ modes.

%This additional mode, particularly its memory contribution, causes a noticeable shift in the strain.  
%However, it's important to note
%Conversely, as the inclination 
%$\Theta$ approaches multiples of $\pi$ the significance of the (2,0) mode progressively diminishes.

In Fig.~\ref{fig:quadrupole_q1_e07}, we present an analogous plot, but for a system with an 
initial EOB eccentricity $e_{\rm EOB}^0=0.7$ and $\langle f_0 \rangle =0.002 M^{-1}$. The results are qualitatively similar to the previous case,
but it is noteworthy that the 2.5PN oscillatory contributions visibly 
break the symmetry of the waveform around periastron. 
While numerical results from perturbation theory suggest that the fluxes—and thus the waveforms—are indeed asymmetric with respect to the periastron 
passage~\cite{Albanesi:2021rby,Placidi:2021rkh,Faggioli:2024ugn}, 
further investigations are needed to assess whether this phenomenology is 
expected in the comparable mass case, or if it is merely an artifact of the 
PN expansion that could be corrected through a proper resummation.
It is important to note that we are utilizing a small eccentricity expansion for the hereditary contributions, and therefore do not expect our results to be accurate a priori at high eccentricities. However, it is worth mentioning that Ref.~\cite{Placidi:2021rkh} demonstrated that eccentricity-expanded tails remain reliable even at high eccentricities, at least in the test-mass scenario.

To assess this issue and, more generally, to asses the reliability of our analytical results, it would be necessary
to perform comparisons with NR
eccentric waveforms. 
However, several factors need to be considered. Firstly, resolving the null memory in NR simulations is not straightforward and typically necessitates Cauchy characteristic 
extraction~\cite{Bishop:1996gt,Reisswig:2009rx,Pollney:2010hs,Mitman:2020pbt}.
There are, however, approximate methods that can be applied to waveforms extracted at finite distances during post-processing~\cite{Mitman:2020bjf}.
Secondly, the numerical waveforms must be mapped to the appropriate BMS frame for a consistent comparison with PN or EOB waveforms~\cite{Mitman:2021xkq,Mitman:2022kwt}. 
Finally, 
mapping the numerical initial data to EOB data for generic orbits is not straightforward, and optimization procedures can introduce biases, see e.g. Sec.~V of Ref.~\cite{Albanesi:2022ywx}. 
For these reasons, EOB/NR comparisons are deferred to future work.
\begin{figure}
  \centering 
    \includegraphics[width=0.48\textwidth]{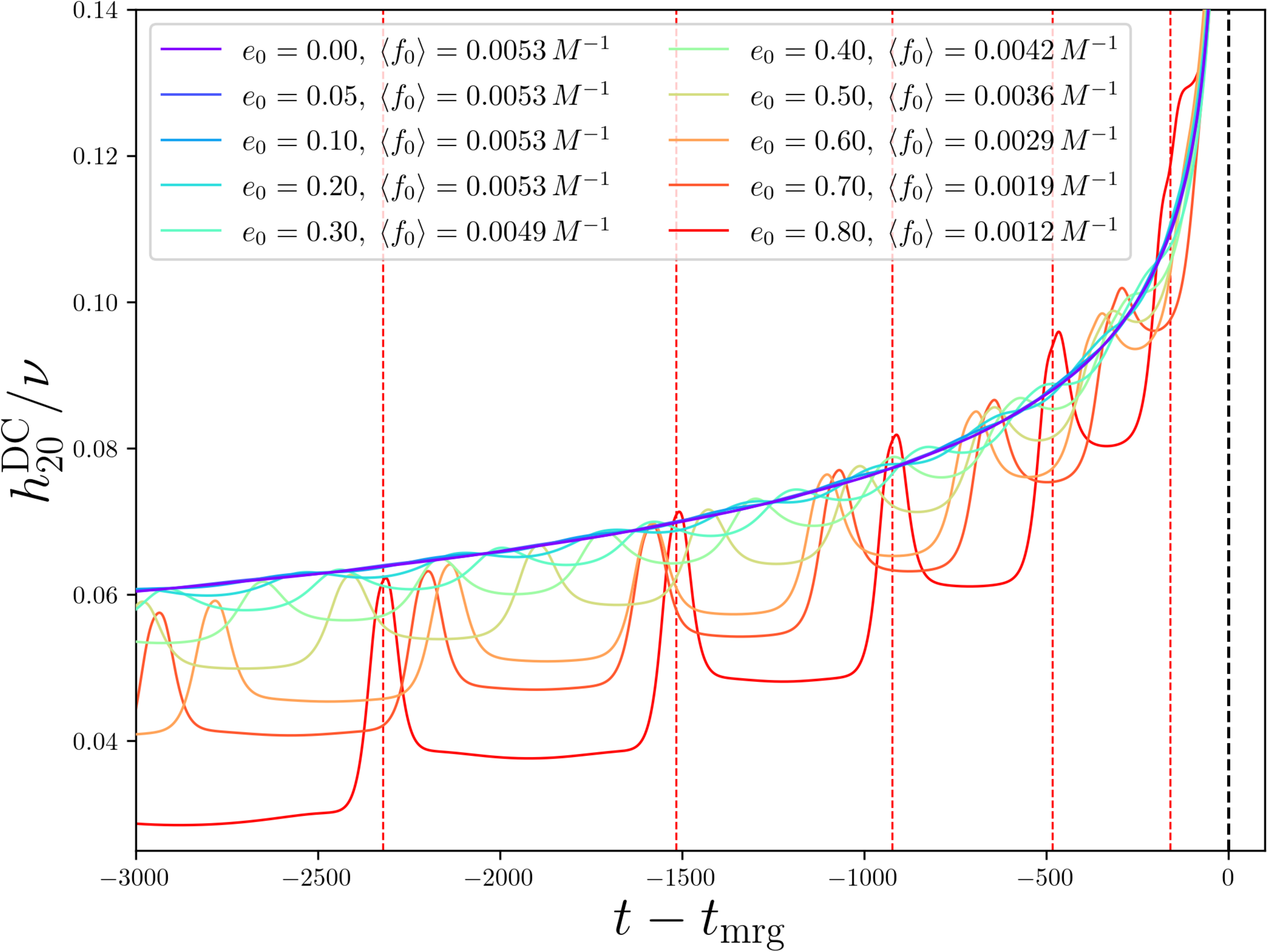}
    \caption{
    DC contribution for different eccentric configurations identified by their initial EOB eccentricity, here simply denoted as $e_0$, and the orbital averaged initial frequency  $\langle f_0 \rangle$.
    The contributions are 
    aligned with respect to the merger of each configuration.
    We also highlight  with dotted vertical lines the periastron 
    passages for the most eccentric system.
    }    
    \label{fig:q1_multi_ecc}
\end{figure} 

To better understand the significance of the displacement memory term across different eccentric systems, we present $h_{20}^{\rm DC}$ in Fig.~\ref{fig:q1_multi_ecc} 
calculated for various eccentric dynamics.  The considered evolutions range from the quasi-circular case (purple) to an initial EOB eccentricity of $e_{\rm EOB}^0=0.8$ (red). The initial orbital 
averaged frequencies $\langle f_0 \rangle$ for each value of 
$e_{\rm EOB}^0$ are selected to ensure that the evolutions last slightly longer than $3000 M$.
We avoid significantly longer evolution times to prevent circularization from occurring before 
the time interval shown in Fig.~\ref{fig:q1_multi_ecc}.
For the most eccentric configuration, we also mark the periastron passages with vertical dotted lines. 
Note that while the eccentricity changes throughout the binary's evolution, 
this aspect is not critical for our qualitative discussion.
All the contributions are aligned with respect to the merger times. 
In the early stages, the contributions from the eccentric configurations are significantly lower than those from the quasi-circular or low-eccentricity systems. As the evolution progresses and the eccentricity decreases, the DC memory contribution at periastron approaches the value seen in quasi-circular cases. However, it remains notably lower during the apastron passages, as anticipated.

\begin{figure*}
  \centering 
    \includegraphics[width=0.48\textwidth]{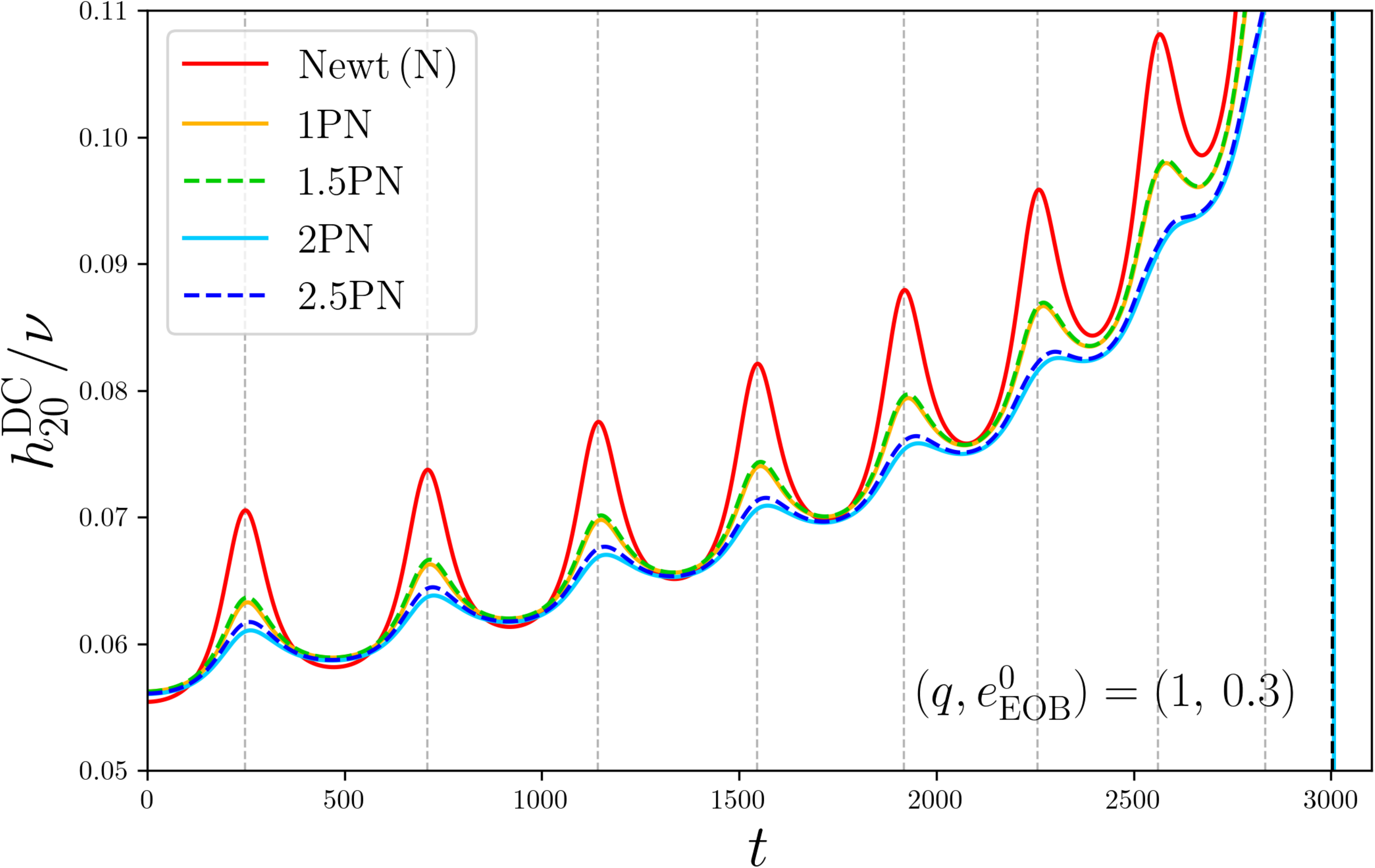}
    \includegraphics[width=0.48\textwidth]{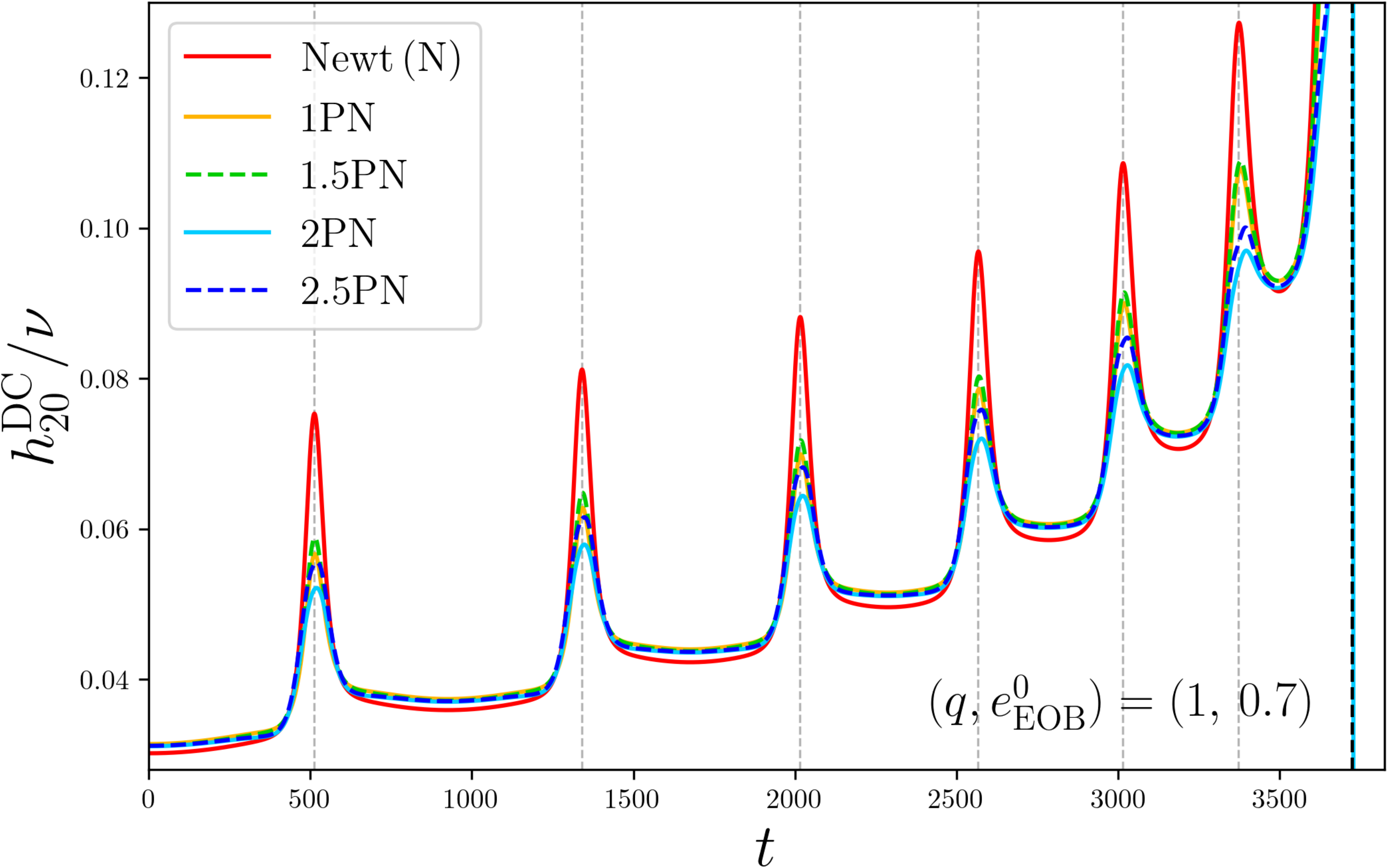}
    \caption{
    DC memory contribution evaluated at different PN orders for the two eccentric configurations considered
    in Fig.~\ref{fig:quadrupole_q1_e03} and~\ref{fig:quadrupole_q1_e07}. Half-integer orders are
    shown with dashed lines, while integer ones with solid lines. The vertical lines mark
    the periastron passages (dotted) and the merger time (dashed). 
    }    
    \label{fig:DC_PN}
\end{figure*} 
Given that our results are based on a PN expansion, it is worthy to assess the contribution of each 
order. We report in Fig.~\ref{fig:DC_PN} the contributions of the memory term at different PN orders 
for the two configurations examined in Fig.~\ref{fig:quadrupole_q1_e03} and~\ref{fig:quadrupole_q1_e07}.
As one can see, the PN orders significantly bring down the solely Newtonian term at the periastron, thus yielding 
a smaller memory contribution. Moreover, the contribution of each term tends to 
diminish as the order increases, indicating convergent behavior. However, the half-integer 
orders (dashed lines) are slightly larger at periastron compared to the preceding 
integer order (solid lines), suggesting that the PN series still exhibits mild oscillations.
The impact of the 2PN and 2.5PN terms on the final result is, however, not negligible, 
suggesting that the inclusion of higher-order corrections or a proper resummation of the 
results could significantly influence the relevance of the DC term.

\begin{figure}
  \centering 
    \includegraphics[width=0.48\textwidth]{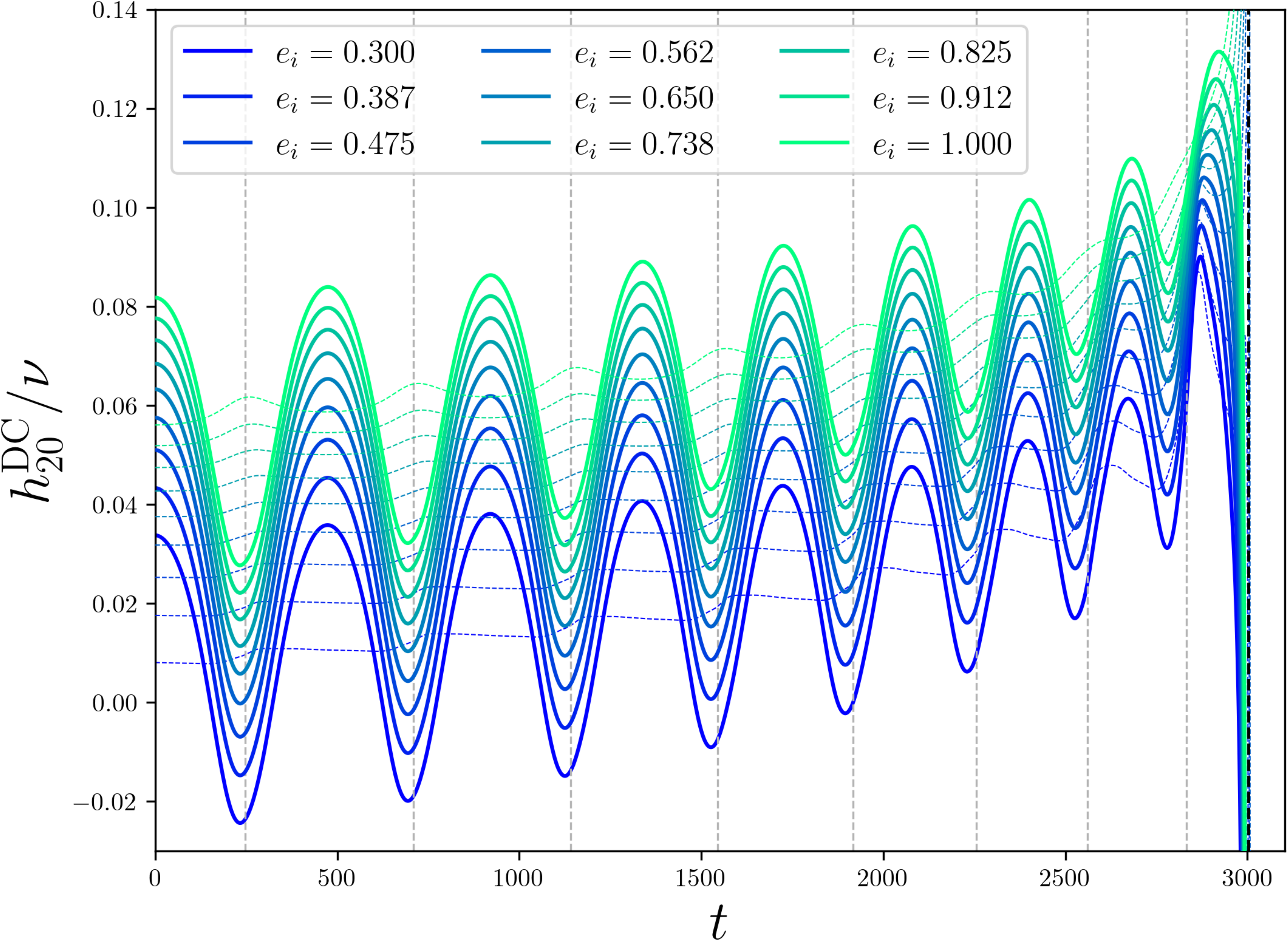}
    \caption{
    Impact of the initial eccentricity $e_i$ 
    for the mildly eccentric system considered in Fig.~\ref{fig:quadrupole_q1_e03}.
    We show the complete (2,0) modes with solid lines
    and the solely DC memory contribution with dashed lines.
    Vertical dotted lines mark periastra passage, while
    the vertical dashed line marks the merger time. 
    }    
    \label{fig:q1_e03_ecci}
\end{figure} 
We conclude this section by briefly discussing the significance of the initial eccentricity
$e_i\equiv e_t(-\infty)$ on the waveform. As previously mentioned, 
this parameter can be interpreted as the eccentricity of the binary at the time of its formation
and should not be confused with the EOB eccentricity
$e_{\rm EOB}^0$ used to generate the initial data for our EOB evolution.
In Fig.~\ref{fig:q1_e03_ecci} we examine the system of
Fig.~\ref{fig:quadrupole_q1_e03} and compute
the (2,0) mode assuming different values for $e_i$ within the range $\in [e_{\rm EOB},1]$.
The contribution of the DC term, represented by dashed lines, increases with higher initial eccentricities. We interpret this growth as an indication that a greater eccentricity at the formation of the binary results in a longer inspiral phase, which consequently allows more time for the displacement memory to accumulate.  As mentioned earlier,
$e_i=1$ would imply that the binary formed through a parabolic encounter.  However,
lower values of $e_i$ are more difficult to interpret, as they would exclude the possibility of formation through a dynamical capture.
For these reasons,
we argue that setting $e_i=1$ as in previous works~\cite{Favata:2011qi},
is the most physically motivated choice, at least for the purposes of this work. 

%%%%%%%%%%%%%%%%%%%%%SECTION%%%%%%%%%%%%%%%%%%%%%%%

\section{Conclusions}
\label{sec:conc}
%In this work, we have developed a 2.5PN-accurate EOB model for the $m=0$ spherical modes, with the crucial inclusion of DC memory effects, for the first time in the EOB literature. 
In this work, we have introduced an EOB model for eccentric orbits 
that incorporates the $m=0$ spherical modes at 2.5PN accuracy, 
including, for the first time in the EOB literature, the crucial contribution of DC memory effects,
which enter in the waveform already at Newtonian level.

Since these memory contributions were previously known only in terms of the QK parametrization of the harmonic coordinates, we have also derived the 2PN mapping between the QK orbital parameters and the EOB phase-space polar variables $(r, \varphi, p_r, p_\varphi)$, which is necessary to compute the EOB-coordinate DC memory with our target 2.5PN accuracy. We recall that, consistent with the original QK expression, the EOB DC memory is provided as an expansion for small eccentricity up to six order and depends on $e_i$, the binary system's eccentricity at formation.

With the updated \DALI{} waveform in hand, we proceeded to study the main properties and impact of the newly incorporated DC memory terms on the waveform. 
We analyzed the memory contribution of the (2,0)
mode in two significant eccentric configurations, 
focusing specifically 
on how incorporating this mode 
into the gravitational strain, modifies the 
observable GW polarizations.
A quantitative assessment of the detectability of the DC contribution 
is left to future work.
Additionally, we examined how the importance of this contribution varies with different eccentricities, and we explored the impact of the initial eccentricity at binary formation, $e_i$,
on the memory contribution during the inspiral phase.

The results here presented are based on a PN expansions, and therefore their reliability 
in the late stages of the evolution should be assessed by means of comparisons with 
numerical relativity computations. However, the absence of public eccentric 
simulations with memory effects leads us to defer this step to future work.
%In conclusion, the results presented in this study are based on a PN expansion and should be validated against numerical simulations. However, performing analytical and numerical comparisons in non-circular cases is challenging, partly due to the difficulty of setting up appropriate initial data for such comparisons, as discussed in Sec.~\ref{sec: properties_DC_mem}. Furthermore, public eccentric NR simulations with null memory are not currently available, so we leave these comparisons for future work.

%%%%%%%%%%%%%%%%%%%%%%%%%%%%%%%%%%%%%%%%%%%%%%%%%
\acknowledgments

We are grateful to A. Nagar for valuable discussions and insightful comments on the manuscript. We thank K. Mitman for useful insights into the current state of NR simulations with memory effects. S.A. and A.P. are also thankful to A. Ramberti and D. Maggioli for continuous support during the final stages of this work.
E.G.,~M.O.,~and~A.P.~acknowledge financial support from the Italian Ministry of University and Research (MUR) through the program “Dipartimenti di Eccellenza 2018-2022” (Grant SUPER-C). E.G.~ and M.O.~also acknowledge support from ``Fondo di Ricerca d'Ateneo" 2021 (MEGA) and 2023 (GraMB) of the University of Perugia. M.O.~acknowledges support from the Italian Ministry of University and Research (MUR) via the PRIN 2022ZHYFA2, GRavitational wavEform models for coalescing compAct binaries with eccenTricity (GREAT).  
S.A.~acknowledges support from the Deutsche Forschungsgemeinschaft (DFG) project ``GROOVHY'' (BE 6301/5-1 Projektnummer: 523180871).
%%%%%%%%%%%%%%%%%%%%%%%%%%%%%%%%%%%%%%%%%%%%%%%%%

%%%%%%%%%%%%%%%%%%%%%%%%%%%%%%%%%%%%%%%%%%%%%%%%%
\appendix
\section{2PN-accurate transformations between QK parametrization and EOB coordinates}
\label{App:transformation_QK_to_EOB_2PN}

In Sec.~\ref{sec:from_QK_parameters_to_EOB_coord}, we obtained the coordinates transformations from QK coordinates to EOB coordinates, with 2PN accuracy. In this Appendix, we summarize all the contributions of the PN expansion of $x$, and $e_t$, beyond the Newtonian order.

For $e_t$, we find
\begin{widetext}
\begin{subequations}
    \begin{align}
    &e_{\rm N}=\,\sqrt{ p_r^2\, p_\varphi^2\, +\Bigl(-1\, +p_\varphi^2\, u\Bigr)^2}, \\ \notag
    &e_{\rm 1PN}=\, -\frac{1}{\sqrt{p_r^2 p_\varphi^2+(-1+p_\varphi^2 u)^2}}\Bigl[u \Bigl(-1+p_\varphi^2 u\Bigr) \Bigl(p_\varphi^2 u (7-3 \nu )+p_\varphi^4 u^2 (-2+\nu ) +2 (-1+\nu
   )\Bigr)\\ 
    & \quad+p_r^4 p_\varphi^2 (-2+\nu )+p_r^2 \Bigl(-1+2 p_\varphi^2 u (5-2 \nu )+2 p_\varphi^4 u^2 (-2+\nu )+\nu \Bigr)\Bigr]\,, \\
   \notag
   &e_{\rm 2PN}=\, -\frac{3\,(-5+2 \nu )}{2
   p_\varphi}\sqrt{-p_r^2+u \bigl(2-p_\varphi^2 u\bigr)} \Bigl[p_r^2+u \Bigl(-2+p_\varphi^2
   u\Bigr)\Bigr] \Bigl[p_r^2 p_\varphi^2+\Bigl(-1+p_\varphi^2 u\Bigr)^2\Bigr] \\ \notag
   & \quad+\frac{1}{2  p_\varphi^2 \bigl(p_r^2 p_\varphi^2+(-1+p_\varphi^2 u)^2\bigr)^{3/2}}\Bigl[p_r^8 p_\varphi^6 \Bigl(8-3 \nu +2 \nu ^2\Bigr) +p_r^6 p_\varphi^4
   \Bigl(9+8 \nu +4 \nu ^2\\ \notag
   & \quad+4 p_\varphi^4 u^2 \bigl(8-3 \nu +2 \nu ^2\bigr)-4 p_\varphi^2 u \bigl(22-9 \nu +4
   \nu ^2\bigr)\Bigr) +u \Bigl(-1+p_\varphi^2 u\Bigr)^4 \Bigl(16-28 \nu+p_\varphi^6 u^3 \bigl(8-3 \nu +2 \nu^2\bigr)\\ \notag
   & \quad-2 p_\varphi^4 u^2 \bigl(22-9 \nu +4 \nu ^2\bigr)+2 p_\varphi^2 u \bigl(12+\nu +4 \nu
   ^2\bigr)\Bigr)+p_r^4 p_\varphi^2 \Bigl(-8+25 \nu +2 \nu ^2+6 p_\varphi^8 u^4 \bigl(8-3 \nu +2 \nu
   ^2\bigr) \\ \notag
   & \quad+15 p_\varphi^4 u^2 \bigl(21-6 \nu +4 \nu ^2\bigr)-6 p_\varphi^6 u^3 \bigl(42-17 \nu +8 \nu
   ^2\bigr)-2 p_\varphi^2 u \bigl(45+14 \nu +12 \nu ^2\bigr)\Bigr)\\ \notag
   & \quad+p_r^2 \Bigl(-8+14 \nu -48 p_\varphi^{10} u^5 \bigl(5-2 \nu +\nu ^2\bigr)+4 p_\varphi^{12} u^6 \bigl(8-3 \nu +2 \nu ^2\bigr)-4 p_\varphi^2 u
   \bigl(-6+23 \nu +2 \nu ^2\bigr)\\ \notag
   & \quad+4 p_\varphi^4 u^2 \bigl(42+28 \nu +13 \nu ^2\bigr)+3 p_\varphi^8 u^4
   \bigl(185-62 \nu +36 \nu ^2\bigr)-2 p_\varphi^6 u^3 \bigl(261-34 \nu +56 \nu ^2\bigr)\Bigr)\Bigr]
\end{align}
\end{subequations}
while, coming to $x$, our result is
\begin{subequations}
\begin{align}
    &x_{\rm N}=\, -p_r^2\, +u\Bigl(\, 2\, -p_\varphi^2\, u\Bigr),  \\ \notag
   & x_{\rm 1PN}=\frac{1}{3 p_\varphi^2}\Bigl[2 p_r^2 \Bigl(-3-2 p_\varphi^2 u (-6+\nu )+p_\varphi^4 u^2 (-3+\nu )\Bigr)+p_r^4
   p_\varphi^2 (-3+\nu )+u \Bigl(12+2 p_\varphi^4 u^2 (9-2 \nu )\\ 
   & \quad+p_\varphi^6 u^3 (-3+\nu )+2 p_\varphi^2 u (-9+2 \nu )\Bigr)\Bigr]\,, \\ \notag
   & x_{\rm 2PN}=\frac{ (-5+2 \nu )}{p_\varphi}\sqrt{-p_r^2+u \bigl(2-p_\varphi^2 u\bigr)} \Bigl(p_r^2+u \bigl(-2+p_\varphi^2
   u\bigr)\Bigr)^2+\frac{1}{36 p_\varphi^4}\Bigl[p_r^6 p_\varphi^4 \Bigl(-72+9 \nu -8 \nu ^2\Bigr)\\ \notag
   & \quad-3
   p_r^4 p_\varphi^2 \Bigl(54-20 \nu -2 p_\varphi^2 u \bigl(120-25 \nu +8 \nu^2\bigr)+p_\varphi^4
   u^2 \bigl(72-9 \nu +8 \nu ^2\bigr)\Bigr)-3 p_r^2 \Bigl(198-60 \nu \\ \notag
   & \quad +8 p_\varphi^2 u (-39+10 \nu )-4
   p_\varphi^6 u^3 \bigl(108-21 \nu +8 \nu ^2\bigr)+4 p_\varphi^4 u^2 \bigl(159-17 \nu +8 \nu
   ^2\bigr) +p_\varphi^8 u^4 \bigl(72-9 \nu +8 \nu ^2\bigr)\Bigr)\\ \notag
   & \quad+u \Bigl(1188-360 \nu +6 p_\varphi^2 u
   (-207+70 \nu )+p_\varphi^{10} u^5 \bigl(-72+9 \nu -8 \nu ^2\bigr)+8 p_\varphi^4 u^2 \bigl(171-48 \nu +8 \nu
   ^2\bigr)\\ \notag
   & \quad+6 p_\varphi^8 u^4 \bigl(96-17 \nu +8 \nu ^2\bigr)-6 p_\varphi^6 u^3 \bigl(219-44 \nu +16 \nu
   ^2\bigr)\Bigr)\Bigr]
\end{align}
\end{subequations}
\end{widetext}
%%%%%%%%%%%%%%%%%%%%%SECTION%%%%%%%%%%%%%%%$$%%%%%%%%

\section{2.5PN contributions to \texorpdfstring{$H_{20}$}{H20} beyond the DC memory}
\label{App:2.5PN_h20}
In this appendix we explicitly provide the 2.5PN contributions to the instantaneous, tail, oscillatory memory, and post-adiabatic components of $H_{20}$ [see Eq.\eqref{eq:sphericalmodesConvention}]. We recall that the same components are given for all relevant spherical modes up to 2PN in Ref.~\cite{Placidi:2021rkh} and up to 2.5PN in the Supplementary Material of this paper.

By recasting in EOB coordinates $(p_{r_*},\dot p_{r_*}, u)$ the results of Refs.~\cite{Mishra:2015bqa,Boetzel:2019nfw,Ebersold:2019kdc} as explained in the main text, we find
\begin{widetext}
\begin{align}
    &H^{\rm inst_{\rm 2.5PN}}_{20} = \frac{  \sqrt{2} }{c^5} \nu\left(-\frac{356}{63} u^3 p_{r_*}+\frac{20}{63} u^2 p_{r_*}^3-\frac{232}{63} u p_{r_*}
   \dot{p}_{r_*}\right)\,, \\ \notag
    &H^{\rm tail_{\rm 2.5PN}}_{20} = \frac{\sqrt{2}}{c^5} \bigg\{\left(-3 u^3 p_{r_*}+3 u p_{r_*} \dot{p}_{r_*}-\frac{3 p_{r_*} \dot{p}_{r_*}^2}{u}+\frac{3
   p_{r_*} \dot{p}_{r_*}^3}{u^3}-\frac{3 p_{r_*} \dot{p}_{r_*}^4}{u^5}+\frac{3 p_{r_*}
   \dot{p}_{r_*}^5}{u^7}\right)
   \\ \notag 
   &\quad +\pi \sqrt{u} \bigg[\frac{1}{7} (33-8 \nu ) u^2  p_{r_*}^2-\frac{5}{168} (-29+3 \nu )
   u p_{r_*}^4+\frac{(31777-11023 \nu ) p_{r_*}^6}{26880} \\ \notag 
   &\quad +\frac{1}{14} (-41+4 \nu )u
   \dot{p}_{r_*}+\frac{1}{14} (5-\nu ) p_{r_*}^2 \dot{p}_{r_*}+\frac{(-1233-4 \nu )
   p_{r_*}^4 \dot{p}_{r_*}}{2688 u}+\frac{(-81+47 \nu ) \dot{p}_{r_*}^2}{56
   u} \\ \notag 
   &\quad+\frac{(1-45 \nu ) p_{r_*}^2 \dot{p}_{r_*}^2}{224 u^2}+\frac{(40081-13579 \nu )
   p_{r_*}^4 \dot{p}_{r_*}^2}{1536 u^3}+\frac{(-52-33 \nu ) \dot{p}_{r_*}^3}{112
   u^3} \\ \notag 
   &\quad+\frac{(-233-43 \nu ) p_{r_*}^2 \dot{p}_{r_*}^3}{672 u^4}+\frac{7 (13+\nu )
   \dot{p}_{r_*}^4}{192 u^5}+\frac{(314137-104105 \nu ) p_{r_*}^2 \dot{p}_{r_*}^4}{1792
   u^6} \\ 
   &\quad+\frac{(-965+46 \nu ) \dot{p}_{r_*}^5}{5376 u^7}+\frac{(-10636323+3547909 \nu )
   \dot{p}_{r_*}^6}{107520 u^9}\bigg] \bigg\}\,, \\ \notag 
   &H_{20}^{\rm mem,osc_{\rm 2.5PN}} = \frac{  \sqrt{2} }{c^5} \nu\bigg(-\frac{96}{7} u^3 p_{r_*}-\frac{1210}{63} u^2 p_{r_*}^3+\frac{61}{90} u
   p_{r_*}^5+\frac{793}{21} u p_{r_*} \dot{p}_{r_*}+\frac{11855}{252} p_{r_*}^3
   \dot{p}_{r_*} \\ \notag 
   &\quad-\frac{1327 p_{r_*}^5 \dot{p}_{r_*}}{630 u}-\frac{2045 p_{r_*}
   \dot{p}_{r_*}^2}{21 u}-\frac{10309 p_{r_*}^3 \dot{p}_{r_*}^2}{126 u^2}+\frac{17095
   p_{r_*} \dot{p}_{r_*}^3}{84 u^3}+\frac{7649 p_{r_*}^3 \dot{p}_{r_*}^3}{63
   u^4} \\ 
   &\quad-\frac{30787 p_{r_*} \dot{p}_{r_*}^4}{84 u^5}+\frac{50005 p_{r_*}
   \dot{p}_{r_*}^5}{84 u^7}\bigg)\,,  \\
   \notag 
   &H_{20}^{\rm post-ad_{\rm 2.5PN}}=\frac{\sqrt{2}}{c^5} \nu\bigg(\frac{16}{15} u^3 \nu  p_{r_*}+\frac{14659}{120} u^2 \nu  p_{r_*}^3+\frac{37771}{96} u
   \nu  p_{r_*}^5+\frac{512}{9} u \nu  p_{r_*} \dot{p}_{r_*}+\frac{1121}{9} \nu 
   p_{r_*}^3 \dot{p}_{r_*} \\ \notag 
   &\quad-\frac{1549981 \nu  p_{r_*}^5 \dot{p}_{r_*}}{2400
   u}-\frac{14459 \nu  p_{r_*} \dot{p}_{r_*}^2}{360 u}+\frac{135503 \nu  p_{r_*}^3
   \dot{p}_{r_*}^2}{240 u^2}+\frac{12843 \nu  p_{r_*} \dot{p}_{r_*}^3}{40
   u^3} \\ 
   &\quad+\frac{631316 \nu  p_{r_*}^3 \dot{p}_{r_*}^3}{225 u^4}-\frac{105091 \nu  p_{r_*}
   \dot{p}_{r_*}^4}{288 u^5}+\frac{17411719 \nu  p_{r_*} \dot{p}_{r_*}^5}{36000 u^7}\bigg).
\end{align}
\end{widetext}
%%%%%%%%%%%%%%%%%%%%%SECTION%%%%%%%%%%%%%%%%%%%%%%%

\section{DC memory for \texorpdfstring{$\ell>2$}{l>2}}
\label{App: Higher_Modes_Hlm}
In this appendix we write the 2.5PN-accurate expressions of DC memory contribution to each $m=0$ mode beyond $\ell=2$. 
%Recall that the dominant $\ell=2$ mode scales as $\mathcal{O}(c^{-4})$, as can be seen from the memory integrand in Eq.~\eqref{eq:Ulmem}. The non-oscillatory contributions to the memory accumulate and are enhanced by a factor of $c^5$ due to integration over the past history. For a complete understanding of the $m=0$ contributions to the waveform up to 2.5PN order, it is necessary to compute the modes up to $\ell =8$. This is due to certain selection rules involved in the angular integral of the product of three spin-weighted spherical harmonics, arising from the energy flux present in the integral of Eq.~\eqref{eq:Ulmem}, As explained in the calculation of the memory contributions accurate to 3PN in harmonic coordinates~\cite{Favata:2008yd, Ebersold:2019kdc}. 
For simplicity, we stop here at the leading order in the small eccentricity expansion and point to the Supplementary Material whomever may be interested in the full expressions up to the sixth order in eccentricity. 

We find
\begin{widetext}
\begin{equation}
    H_{40}^{\rm DC}\,=\, -\frac{u}{502 \sqrt{2}}\Biggl(H_{40}^{\rm DC_{\rm N}}\, +u\, H_{40}^{\rm DC_{\rm 1PN} }\,+ u^{3/2}\,H_{40}^{\rm DC_{\rm 1.5PN}}\, +u^2\, H_{40}^{\rm DC_{\rm 2PN} }\,+ u^{5/2}\,H_{40}^{\rm DC_{\rm 2.5PN}}\Biggr),
\end{equation}
\begin{subequations}
    \begin{align}
        &H_{40}^{\rm DC_{\rm N}}\, =\, 1\, -\Biggl( \frac{Z_p}{e_i u^2}\Biggr)^{\frac{12}{19}}, \\ \notag 
        &H_{40}^{\rm DC_{\rm 1PN }}=\,-\frac{180101}{29568}\, +\frac{9193}{352}\nu\, +\Biggl( \frac{Z_p}{e_i u^2}\Biggr)^{\frac{24}{19}}\Biggl(\frac{3920527}{561792}\, -\frac{183995}{6688}\nu\Biggr)\, +\Biggl( \frac{Z_p}{e_i u^2}\Biggr)^{\frac{12}{19}}\Biggl[-\frac{2833}{3192}\, +\frac{53}{38}\nu\, \\
   & \quad+\frac{6}{19 Z_p^4}\Biggl( p_{r_*}^4\, u^6\bigl(\, 3\, -2\nu\bigr)\, -\Dot{p}_{r_*}^4\bigl(\, 3\, +\nu\bigr)\, -3\, p_{r_*}^2\, \Dot{p}_{r_*}^2\, u^3 \nu\Biggr) \Biggr], \\
        &H_{40}^{\rm DC_{\rm 1.5PN}}=\, -\frac{377}{228}\pi\, \Biggl( \frac{Z_p}{e_i u^2}\Biggr)^{\frac{12}{19}} \Biggl[\, 1-\Biggl( \frac{Z_p}{e_i u^2}\Biggr)^{\frac{30}{19}}\Biggr],  \\
   \notag 
        &H_{40}^{\rm DC_{\rm 2PN} }\, =\, \frac{2201411267}{158505983}\, -\frac{13500365}{157248}\nu\, +\frac{1322987}{27456}\nu^2\, +\Biggl( \frac{Z_p}{e_i u^2}\Biggr)^{\frac{24}{19}}\Biggl[\frac{11106852991}{896620032}\,-\frac{364522883}{5337024}\nu\, +\frac{9751735}{120072}\nu^2\, \\ \notag
   & \quad+\frac{1}{1123584\, Z_p^2}\Biggl(\, p_{r_*}^2\, u^3\bigl(-11761581\, +54207794\nu\, +30911160\nu^2\bigr)\, +\Dot{p}_{r_*}^2\, \bigl(117611581\, -42446213\nu\, -15455580\nu^2\bigr)\Biggr)\,  \\ \notag 
   &\quad +\frac{1}{21348096\, Z_p^4}\Biggl(p_{r_*}^2\, \Dot{p}_{r_*}^2\, u^3\bigl(\, 58807905\nu\, -231833700 \nu^2\bigr)\, +p_{r_*}^4\, u^6\, \bigl(\, -58807905\, +271038970\nu\, -154555800\nu^2\bigr)\,\\ \notag
   & \quad +\Dot{p}_{r_*}^4\bigl(\, 58807905\, -212231065\nu\, -77277900 \nu^2\bigr)\Biggl)  \Biggr]\, +\Biggl( \frac{Z_p}{e_i u^2}\Biggr)^{\frac{12}{19}}\Biggl[\, \frac{358353209}{366799104}\, -\frac{259303}{727776}\nu\, -\frac{1495}{5776}\nu^2\,\\ \notag
   & \quad +\frac{1}{722\, Z_p^4} \Biggl(p_{r_*}^4\, \Dot{p}_{r_*}^4\, u^6\bigl(-19188\, +19107\nu\, -1722\nu^2\bigr)\, +p_{r_*}^6\, \Dot{p}_{r_*}^2\, u^9\bigl(-7752\, +11193\nu\, -1344\nu^2\bigr)\, \\ \notag
   & \quad+p_{r_*}^2\, \Dot{p}_{r_*}^6\, u^3\bigl(\, -15960\, +13659\nu\, -900\nu^2\bigr)\, +p_{r_*}^8\, u^{12}\bigl(-210\, +2256\nu\, -372\nu^2\bigr)\, +p_{r_*}^8\bigl(-4314\, +3489\nu\, -150\nu^2\bigr)\Biggl)\,\\ \notag
   & \quad+\frac{1}{10108\, Z_p^4}\Biggl(\, p_{r_*}^4\, \bigl(-8499\, +10523\nu\, +4452\nu^2\bigr)\, +p_{r_*}^4\, u^6\bigl(8499\, -19022\nu\, +8904\nu^2\bigr)\,  \\ 
   &\quad +p_{r_*}^2\,\Dot{p}_{r_*}^2\, u^3 \bigl(-8499\nu\, +13356\nu^2\bigr)\Biggr) \Biggr] ,  \\
   \notag 
        &H_{40}^{\rm DC_{\rm 2.5PN} }= \, \pi\Biggl\{\, -\frac{13565}{1232}\,+\frac{1365}{308}\nu\, +\Biggl( \frac{Z_p}{e_i u^2}\Biggr)^{\frac{24}{19}}\Biggl(\frac{147803879}{64044288}-\frac{69366115}{762432}\nu\Biggr)\,+\Biggl( \frac{Z_p}{e_i u^2}\Biggr)^{\frac{23}{19}}\Biggl(\, -\frac{473166857}{29111040}\,\\ \notag
   &\quad  +\frac{1255597433}{26685120}\nu \Biggr)\, +\Biggl( \frac{Z_p}{e_i u^2}\Biggr)^{\frac{12}{19}}\Biggl[\frac{3763903}{7277760}\, +\frac{3427243}{606480}\nu\, +\frac{1}{722\, Z_p^4}\Biggl(\, -p_{r_*}^4\bigl(\, 1131\, +377\nu\bigr)\, +p_{r_*}^4\, u^6\bigl(1131\, -754\nu\bigr)\,   \\   \notag
        & \quad -1131\, p_{r_*}^2\Dot{p}_{r_*}^2\, u^3\nu\Biggr) \Biggr]+\Biggl( \frac{Z_p}{e_i u^2}\Biggr)^{\frac{30}{19}}\Biggl[\frac{5340205}{1455552}\, -\frac{99905}{17328}\nu\,+ \frac{1}{1444\, Z_p^4}\Biggl(\, \Dot{p}_{r_*}^4\bigl(\, 5655\, +1855\nu\bigr)\, \\
   &\quad  +5655\, p_{r_*}^2\, \Dot{p}_{r_*}^2\, u^3\, \nu\,+p_{r_*}^4\, u^6\bigl(\,-5655\, +3770\nu\bigr)\Biggr) \Biggr]\Biggr\} \\ \notag    \end{align}
\end{subequations}
\begin{equation}
    H_{60}^{\rm DC}\,=\, \frac{4195\, u^2}{1419264 \sqrt{273}}\Biggl( H_{60}^{\rm DC_{\rm 1PN} }\, +u\, H_{60}^{\rm DC_{\rm 2PN} }\,+ u^{3/2}\,H_{60}^{\rm DC_{\rm 2.5PN} }\Biggr),
\end{equation}
\begin{subequations}
    \begin{align}
        &H_{60}^{\rm DC_{\rm 1PN} }=\Biggl(\, 1\, -\frac{3612}{839}\nu\Biggr)\Biggl[\, 1\, -\Biggl( \frac{Z_p}{e_i u^2}\Biggr)^{\frac{24}{19}}\Biggr],\\
          &H_{60}^{\rm DC_{\rm 2PN} }=\, -\frac{455661561}{6342840}\, +\frac{34364}{839}\nu\, -\frac{50526}{839}\nu^2\, + \Biggl( \frac{Z_p}{e_i u^2}\Biggr)^{\frac{36}{19}} \Biggl( \frac{1081489489}{120513960}\, -\frac{819202}{159411}\nu\, +\frac{1151430}{15941}\nu^2\Biggr)\, \\ \notag
   &\quad +\Biggl( \frac{Z_p}{e_i u^2}\Biggr)^{\frac{24}{19}}\Biggl[\, -\frac{2833}{1596}\, +\frac{166286}{15941}\nu\, -\frac{191436}{15941}\nu^2\,+\frac{1}{31882\, Z_p^4}\Biggl(\, \Dot{p}_{r_*}^4\, \bigl(\, -12585\, +49985\nu\, +18060\nu^2\bigr)\,\\ \notag
   &\quad  +p_{r_*}^4\, u^6\bigl(\, 12585\, -62570\nu\,+36120\nu^2\bigr)\,+p_{r_*}^2\Dot{p}_{r_*}^2\, u^3\bigl(\,-12585 \nu\, +54180\nu^2\bigr)\Biggr)  \Biggr]  \\ \notag 
           &H_{60}^{\rm DC_{\rm 2.5PN} }= \, \pi\Biggl[\, \frac{1248}{839}\, -\frac{4992}{839}\nu\, +\Biggl( \frac{Z_p}{e_i u^2}\Biggr)^{\frac{24}{19}}\Biggl(-\frac{377}{144}\, +\frac{226954}{15941}\nu\Biggr)\,+\Biggl( \frac{Z_p}{e_i u^2}\Biggr)^{\frac{23}{19}} \Biggl(\frac{174031}{95646}\, -\frac{132106}{15941}\nu\Biggr) \Biggr] \\ \notag
    \end{align}
\end{subequations}
\begin{equation}
    H_{80}^{\rm DC}=\, -\frac{75601}{213497856 \sqrt{119}}\, u^3\Biggl(1\, -\frac{452070}{75601}\nu\, +\frac{733320}{75601}\nu^2\Biggr)\Biggl[1\, -\Biggl( \frac{Z_p}{e_i u^2}\Biggr)^{\frac{36}{19}}\Biggr].
\end{equation}
\end{widetext}
We emphasize that the $\ell=8$ mode, which normally would not appear before the 3PN order, enters the 2.5PN-accurate waveform only with its 2PN DC term, due to the accumulation effects discussed in the main text above Eq.~\eqref{eq:Ulmem}.

%%%%%%%%%%%%%%%%%%%% REFERENCES %%%%%%%%%%%%%%%%%%
\newpage
\bibliography{Bibliografy} 

\end{document}